\documentclass[twocolumn,prb,showpacs,superscriptaddress,floatfix,amsmath,amssymb,amsfonts,aps]{revtex4-1}
\usepackage{graphicx}
\pdfoutput=1
\usepackage{multirow}
\usepackage{color}
\usepackage[colorlinks,bookmarks=false,citecolor=blue,linkcolor=red,urlcolor=blue]{hyperref}
\newcommand{\cuf}{CuF$_2$}
\newcommand{\cubr}{CuBr$_2$}
\newcommand{\cucl}{CuCl$_2$}

\newcommand{\cuxx}{CuX$_2$ (X\,=\,F, Cl, Br)}
\newcommand{\AFM}{\text{AFM}}
\newcommand{\FM}{\text{FM}}
\newcommand{\eff}{\text{eff}}
\begin{document}

\title{Magnetism of CuX$_2$ frustrated chains (X = F, Cl, Br): the role of covalency}

\author{S. Lebernegg}
\email{stefan.l@sbg.at}
\affiliation{Max Planck Institute for Chemical Physics of Solids, 01187 Dresden, Germany}

\author{M. Schmitt}
\affiliation{Max Planck Institute for Chemical Physics of Solids, 01187 Dresden, Germany}

\author{A. A. Tsirlin}
\affiliation{Max Planck Institute for Chemical Physics of Solids, 01187 Dresden, Germany}
\affiliation{National Institute of Chemical Physics and Biophysics, 12618 Tallinn, Estonia}

\author{O. Janson}
\affiliation{Max Planck Institute for Chemical Physics of Solids, 01187 Dresden, Germany}

\author{H. Rosner}
\email{rosner@cpfs.mpg.de}
\affiliation{Max Planck Institute for Chemical Physics of Solids, 01187 Dresden, Germany}

\date{\today}

\begin{abstract}
Periodic and cluster density-functional theory (DFT) calculations, including
DFT+$U$ and \mbox{hybrid} functionals, are applied to study magnetostructural
correlations in spin-1/2 frustrated chain \mbox{compounds} CuX$_2$: CuCl$_2$,
CuBr$_2$, and a fictitious chain structure of CuF$_2$.  The nearest-neighbor
and second-neighbor exchange integrals, $J_1$ and $J_2$, are evaluated as a
function of the Cu--X--Cu bridging angle $\theta$ in the physically relevant
range 80--110$^{\circ}$. In the ionic CuF$_2$,  $J_1$ is \mbox{ferromagnetic} for
$\theta\leq100^{\circ}$. For larger angles, the antiferromagnetic superexchange
contribution becomes \mbox{dominant,} in accord with the Goodenough-Kanamori-Anderson rules.
However, both CuCl$_2$ and CuBr$_2$ \mbox{feature} ferromagnetic $J_1$ in the whole
angular range studied.  This surprising behavior is ascribed to the increased
covalency in the Cl and Br compounds, which amplifies the contribution from Hund's \mbox{exchange} on
the ligand atoms and renders $J_1$ ferromagnetic.  At the same time, the larger
\mbox{spatial} extent of X orbitals enhances the antiferromagnetic $J_2$, which is
realized via the long-range \mbox{Cu--X--X--Cu} paths.  
Both, periodic and cluster approaches supply a consistent description of the magnetic 
\mbox{behavior} which is in good agreement with the experimental
data for CuCl$_2$ and CuBr$_2$. Thus, owing to their simplicity, cluster calculations 
have excellent potential to study magnetic \mbox{correlations} in more involved spin lattices and facilitate application of quantum-chemical methods.

\end{abstract}


\maketitle

\section{Introduction}
Copper compounds have been extensively studied as spin-$\frac12$ quantum
magnets, material prototypes of quantum spin models. While local properties of
these compounds are usually similar and involve nearly isotropic Heisenberg
spins, the variability of the magnetic behavior stems from the unique
structural diversity. Depending on the particular arrangement of the magnetic
Cu$^{2+}$ atoms and their ligands in the crystal structure, different spin
lattices can be formed. Presently, experimental examples for many of simple
lattice geometries, including the uniform
chain,\cite{stone2003,HC_KCuF3_INS_5_LL} square
lattice,\cite{ronnow1999,*ronnow2001,tsyrulin2009} Shastry-Sutherland lattice
of orthogonal spin dimers,\cite{miyahara2003,takigawa2010} and kagom\'e
lattice,\cite{mendels2010} are available and actively studied. Some of the
copper compounds feature more complex spin
lattices\cite{ruegg2007,mentre2009,*tsirlin2010,janson2012} that have not been
anticipated in theoretical studies, yet trigger the theoretical
research\cite{laflorencie2011,lavarelo2011} once relevant material prototypes
are available.

Owing to the competition between ferromagnetic (FM) and antiferromagnetic (AFM) contributions 
to the exchange couplings, compounds of particular interest are those with M-X-M bridging angles 
close to 90$^{\circ}$, with M being a transition metal and X being a ligand. Such geometries 
are realized in the quasi-1D cuprates featuring chains of edge-sharing CuO$_4$ plaquettes, 
which represent a simple example of low-dimensional spin-1/2 magnetic materials. Independent 
of the sign of the nearest-neighbor (NN) coupling $J_1$, its competition with  
the sizeable AFM next-nearest-neighbor (NNN) coupling $J_2$ leads to magnetic frustration. 
Depending on the ratio $J_2/J_1$, such compounds exhibit exotic magnetic behavior like helical 
order,\cite{FHC_LiCuVO4_ENS_spin_supercurrents} spin-Peierls transition\cite{FHC_CuGeO3_spin-Peierls} 
or quantum critical behavior.\cite{FHC_Li2ZrCuO4} The difficulties in the 
microscopic description of such compounds originate from 
ambiguities\footnote{The same value of specific physical properties such as propagation vectors 
or spin gaps can be realized in different parts of the phase-diagram. 
To illustrate this effect, we consider an AFM $J_1$
and $J_2$\,=\,0. In this case, the model is reduced to a uniform
Heisenberg chain. In the other limit, $J_1$\,=\,0 with an AFM $J_2$, the
system reduces to two decoupled uniform Heisenberg chains. Although the
$J_2/J_1$ ratios are very different for these two cases, the physics is
the same.} in the experimental estimates of the ratio $J_2/J_1$, leading to controversial 
modeling of the magnetic structure.\cite{FHC_LiCu2O2_chiT_CpT_NS_wrong_model_paper, 
*FHC_LiCu2O2_chiT_CpT_NS_wrong_model_comment, *FHC_LiCu2O2_chiT_CpT_NS_wrong_model_reply} 
Thus, the combination of different sets of experimental data with a careful theoretical analysis 
of the individual exchange pathways is of crucial importance for obtaining a precise microscopic magnetic model.

However, the search for new quantum magnets, as well as the work on existing materials,
require not only the ability to estimate the couplings but also a solid understanding of 
the nexus between crystallographic features of
the material and ensuing magnetic couplings. The Goodenough-Kanamori-Anderson
(GKA)\cite{GKA_1,*GKA_2,*GKA_3} rules are a generic and well-established
paradigm that prescribes FM couplings for bridging angles
close to 90$^{\circ}$ and AFM couplings else, where the bridging angle refers to the 
M--X--M pathway. In Cu$^{2+}$ oxides, generally the
GKA rules successfully explain the crossover between the FM and AFM
interactions for Cu--O--Cu angles close to 90$^{\circ}$. The boundary between
the FM and AFM regimes is usually within the range of $95-98^{\circ}$,\cite{braden} 
but may considerably be altered by side groups and distortions.\cite{gk, ruiz97_2, leb2}

In addition to Cu$^{2+}$ oxides, the systems of interest include copper
halides,\cite{coldea1997,*coldea2001,ruegg2003}
carbodiimides,\cite{zorko2011,*tsirlin2012} and other compound families.
Although microscopic arguments behind the GKA rules should be also applicable
to these non-oxide materials, the critical angles separating the FM and AFM
regimes, as well as the role of the ligand in general, are still little
explored. Moreover, the low number of experimentally studied compounds impedes
a comprehensive experimental analysis available for oxides. 

More and more, density functional theory (DFT) electronic structure calculations complement 
experimental studies and deliver accurate estimates of magnetic couplings.\cite{leb_a2cup2o7,dioptase,azurite,licu2o2_whangbo,cav4o9_pick,WF_Ku_2002} 
They are especially well suited for the study of magnetostructural correlations, as both real and fictitious
crystal structures can be considered in a calculation. However, in a periodic
structure the effect of a single geometrical parameter is often difficult to
elucidate, because different geometrical features are intertwined and evolve
simultaneously upon the variation of an atomic position. Geometrical effects on the local magnetic coupling
are better discerned in cluster models that represent a small group of
magnetic atoms and, ideally, a single exchange pathway. Additional advantages of 
cluster models, owing to their low number of correlated atoms, are lower computational 
costs and, most important, their potential for the application of parameter-free 
wavefunction-based computational methods, i.e. in a strict sense $ab$ $initio$ 
calculations. By contrast, presently available band-structure methods for calculating 
strongly correlated compounds rely on empirical parameters and corrections where their 
choice is in general not unambiguous.\cite{dcc1,dcc2}

There have been several attempts to describe the local properties of 
solids with clusters especially in combination with $ab$ $initio$ quantum-chemical 
methods.\cite{munoz2002,munoz2005,caballol2010,hozoi_2011,paulus_2008} However, the construction of clusters is far from being trivial. 
On one side, to make the calculations computationally feasible, the number of quantum mechanically treated atoms has to be kept as small as possible.  
On the other side, accurate results require that these atoms experience the "true" crystal potential. 
Usually, this is achieved by embedding the cluster into a cloud of 
point charges\cite{paulus_2012,hozoi_2011} and so called total ion potentials.\cite{TIP_winter89,graaf1}
But even for involved embeddings it was demonstrated, that the choice of the cluster 
may have significant effects on the results of the calculations and, thus, size-convergence has to be checked thoroughly.\cite{graaf1}

Here, we study the effect of geometrical parameters on the magnetic exchange in
Cu$^{2+}$ halides. The family of halogen atoms spans a wide range of
electronegativities, from the ultimately electronegative fluorine, forming
strongly ionic Cu--F bonds, to chlorine and bromine that produce largely
covalent compounds with Cu$^{2+}$.\cite{sawatzky1981} Presently, we do not
consider iodine because no Cu$^{2+}$ iodides have been reported. In our
modeling, we use the simplest possible periodic crystal structure of a CuX$_2$
chain that enables the variation of the Cu--X--Cu bridging angle in a broad
range. We further perform a comparative analysis for clusters and additionally
consider the problem of long-range couplings. The evaluation of such couplings
requires larger clusters, thus posing a difficulty for the cluster approach.
The observed trends for the magnetic exchange as a function of the bridging
angle are analyzed from the microscopic viewpoint, and reveal the crucial role
of covalency that underlies salient differences between the ionic Cu$^{2+}$
fluorides and largely covalent chlorides and bromides. 

On the experimental side, the compounds and crystal structures under
consideration are relevant to the CuCl$_2$ and CuBr$_2$ materials that show
interesting examples of frustrated Heisenberg
chains.\cite{banks2009,FHC_CuCl2_DFT_chiT_simul_TMRG} At low temperatures,
these halides form helical magnetic structures and demonstrate improper
ferroelectricity along with the strong magnetoelectric
coupling.\cite{seki2010,zhao2012}

The paper is organized as follows. In section II, the applied theoretical
methods are presented. In the third section, the crystal structures of the
CuX$_2$ compounds are described and compared. In section IV, the results of
periodic and cluster calculations are discussed and compared. Finally, the
discussion, summary, and a short outlook are given in section V.

\section{Methods}
The electronic structures of clusters and periodic systems were calculated with
the full-potential local-orbital code \textsc{fplo9.00-34}.\cite{fplo} For the
scalar-relativistic calculations within the local density approximation (LDA), the Perdew-Wang
parameterization\cite{pw92} of the exchange-correlation potential was used together 
with a well converged mesh of up to 12$\times$12$\times$12 k-points for the periodic models.

The effects of strong electronic correlations were considered by mapping the
LDA bands onto an effective tight-binding (TB) model. The transfer integrals
$t_i$ of the TB-model are evaluated as nondiagonal elements between Wannier
functions (WFs). For the clusters, the transfer integral corresponds to half of the 
energy difference of the magnetic orbitals.\cite{hth75} These transfer integrals $t_i$ are further introduced into the
half-filled single-band Hubbard model $\hat{H}=\hat{H}_{TB}+U_{\text{eff}}\sum_{i}\hat{n}_{i\uparrow}\hat{n}_{i\downarrow}$ that is eventually reduced to the
Heisenberg model for low-energy excitations,
\begin{equation}
\hat{H}=\sum_{\left\langle ij\right\rangle}J_{ij}\hat{S_{i}}\cdot\hat{S_{j}}
\end{equation}
 The reduction is well-justified in
the strongly correlated limit $t_i\ll U_{\eff}$, where $U_{\eff}$ is the
effective on-site Coulomb repulsion, which exceeds $t_i$ by at least an order
of magnitude (see Table~\ref{T_tJ}). This procedure yields AFM contributions to the exchange evaluated
as $J_i^{\AFM}=4t_i^2/U_{\eff}$.

Alternatively, the full exchange couplings $J_i$, comprising FM and AFM contributions, can be derived from total energies
of collinear magnetic arrangements evaluated in spin-polarized supercell calculations\footnote{A unit cell quadrupled along the $b$ axis with P2/m symmetry and five Cu-sites defines the supercell. Three different arrangements of the spins localized on the Cu-sites were sufficient to calculate all presented $J_i$: two with the spin on one Cu-site flipped and one with the spins on two Cu-sites flipped (see supplemental material\cite{supp}). This defines a system of linear equations of the type $E_s=\epsilon_0+a_s\cdot J_1+b_s\cdot J_2$ which can easily be solved. $E_s$ is the total energy of spin arrangement $s$, $\epsilon_0$ is a constant and $a_s$ and $b_s$ describe how often a certain coupling is effectively contained in the supercell.}
within the mean-field density functional theory (DFT)+$U$ formalism. 
We use a local spin-density approximation (LSDA)+$U$ scheme in combination with a unit cell quadrupled along the $b$ axis and a $k$-mesh of 64 points. 
The on-site repulsion and exchange amount to $U_d$\,=\,7$\pm$0.5\,eV and $J_d$\,=\,1\,eV,
respectively. The same $U_d$ value is chosen for all \cuxx\ compounds to facilitate a 
comparison of the magnetic behavior. In section~\ref{sec:luvsb3}, however, it will be 
shown that $U_d$ has in fact no qualitative effect on the magnetic couplings of the \cuxx\ compounds.  
We applied the around mean field (AMF) as well as the fully localized limit (FLL) 
double counting corrections where both types where found to supply similar results. Thus, following the earlier studies of Cu$^{2+}$
compounds,\cite{FHC_CuCl2_DFT_chiT_simul_TMRG,leb_a2cup2o7,dioptase} the presented results are obtained within the AMF scheme.

For the clusters we used, in addition to the LSDA+$U$ method, the B3LYP hybrid
functional\cite{b3lyp} with a 6-311G basis
set. The B3LYP calculations were performed within the \textsc{gaussian09} code.\cite{g09} The free
parameter $\alpha$, indicating the admixture of exact exchange, was varied
in the range between 0.15 and 0.25 to investigate its influence on the calculated exchange couplings.

\section{Crystal structures}
The copper CuX$_2$ dihalides feature isolated chains of edge-sharing CuX$_4$
plaquettes.\cite{supp} The chains of this type are the central building block of many well-studied
cuprates such as CuGeO$_3$ (Ref.~\onlinecite{FHC_CuGeO3_spin-Peierls}),
Li$_2$ZrCuO$_4$ (Ref.~\onlinecite{FHC_Li2ZrCuO4}), and Li$_2$CuO$_2$
CuX$_2$ halides are charge neutral, which makes them especially well suited for the
modeling within the cluster approach.

\begin{figure}[tbp]
\includegraphics[width=8.6cm]{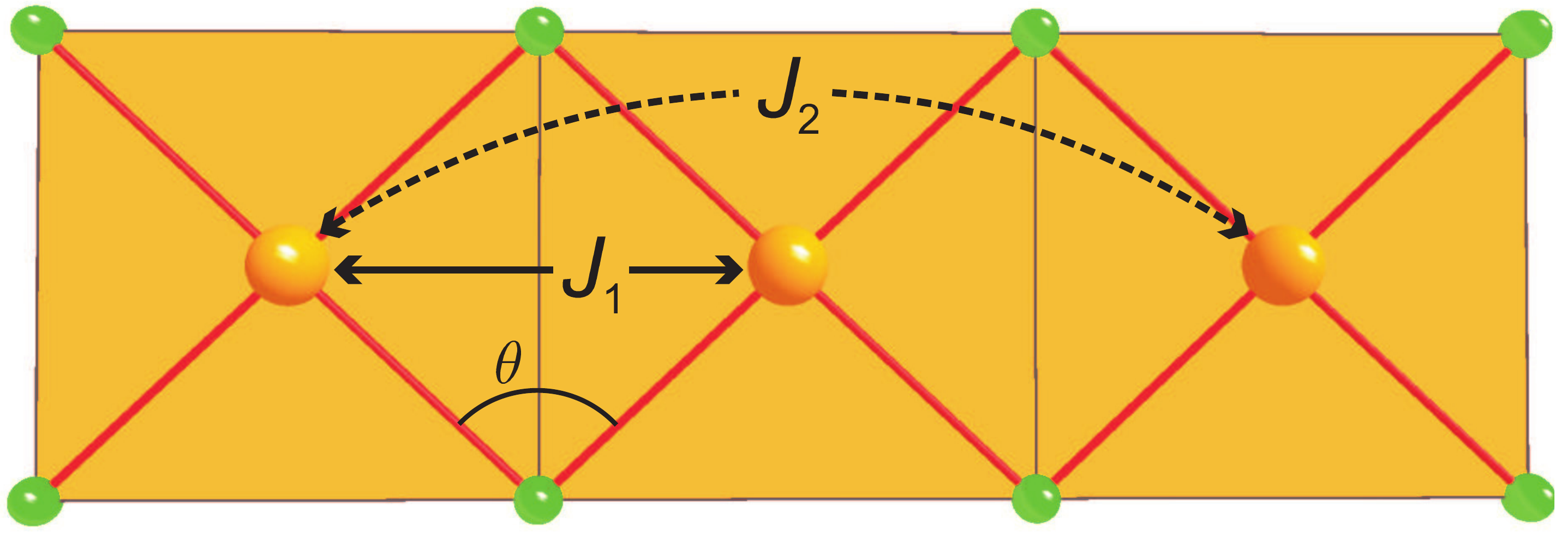}
\caption{\label{F-str}(Color online) Edge-sharing CuX$_4$-plaquettes forming the magnetic chains
in the CuX$_2$ compounds. The chains, running along [010] are flat and lie in the $ab$ plane.  The stacking of the planes is accompanied by a shift to match the
monoclinic angle.\cite{supp} The arrows indicate the nearest-neighbor and
next-nearest-neighbor interaction pathways, and $\theta$ denotes the Cu-X-Cu bridging
angle.}
\end{figure}

\cubr\ crystallizes in the monoclinic space group $C2/m$ with
$a$\,=\,14.728\,\r{A}, $b$\,=\,5.698\,\r{A} and $c$\,=\,8.067\,\r{A}, and
$\beta$\,=\,115.15$^{\circ}$ at room temperature.\cite{oeckler} The planar
chains of edge-sharing CuBr$_{4}$ plaquettes run along the $b$-axis
(Fig.~\ref{F-str}). The Cu--Br--Cu bridging angle $\theta$ amounts to
92.0$^{\circ}$, the Cu--Br distance is 2.41\,\r{A}, while the distances between
the neighboring chains amount to $d_{\parallel}$\,=\,3.82\,\r{A} and
$d_{\perp}$\,=\,3.15\,\r{A} in the direction parallel to $c$ and perpendicular
to the plaquette plane, respectively.

CuCl$_2$ is isostructural to \cubr\ with the Cu--Cl distance of 2.26\,\r{A} and 
$\angle$(Cu--Cl--Cu)\,=\,93.6$^{\circ}$.\cite{cucl2_str} The interchain
separations amount to $d_{\parallel}$\,=\,3.73\,\r{A} and
$d_{\perp}$\,=\,2.96\,\r{A} along the $c$ and $a$ directions, respectively.

\cuf\ features a two-dimensional distorted version of the rutile structure,
with corner-sharing CuF$_4$ plaquettes forming a buckled square
lattice.\cite{cuf2_str} This atomic arrangement is very different from the
chain structures of \cucl\ and \cubr. For the sake of comparison with other
Cu$^{2+}$ halides, we constructed a fictitious one-dimensional structure of
CuF$_2$. The Cu--F distance of 1.91\,\r{A} was chosen to match the respective
average bond length in the real \cuf\ compound. The corresponding bridging angle,
yielding a minimum in total energy, was determined to be 102$^{\circ}$.\footnote{The 
bridging angle minimizing the total energy for the given Cu-F bonding distance was 
estimated by a series of LDA-calculations for bridging angles between 70$^{\circ}$ 
and 120$^{\circ}$.} Although this crystal structure remains hypothetical, it is 
likely metastable and could be formed in CuF$_2$ under a strong tensile strain on 
an appropriate substrate.

\section{Results}

\subsection{Band structure calculations}
\begin{figure}[tbp]
\includegraphics[width=8.6cm]{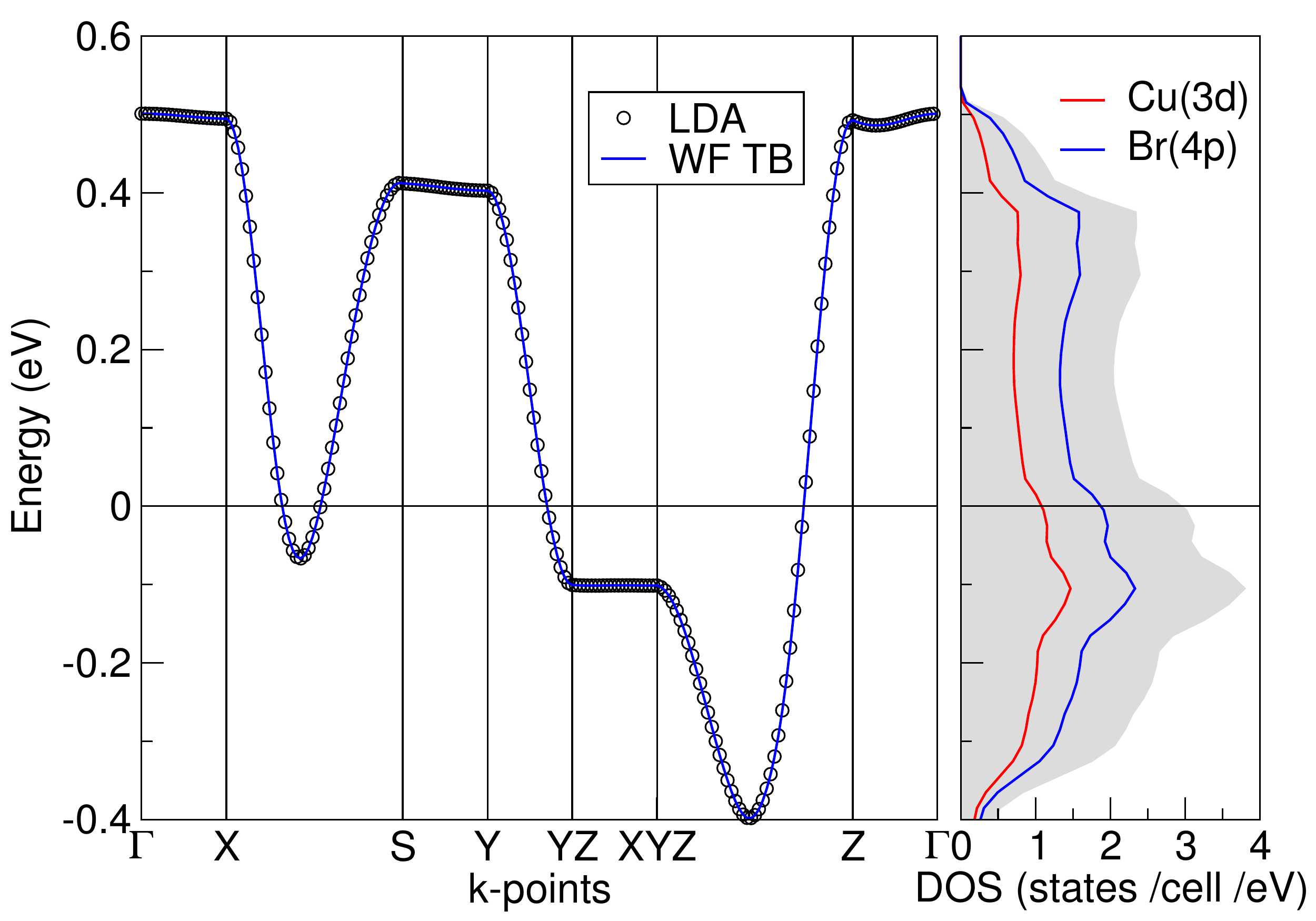}
\caption{\label{bands}(Color online) Comparison of the calculated LDA
band structure of \cubr\ and the band derived from a fit using an effective
one-band tight-binding model based on Cu-centered Wannier functions (WF TB).
The right plot shows the total density of states (DOS) together with the
partial DOS of Cu(3$d$) and Br(4$p$) states. The Fermi level is at zero energy.
Notation of $k$-points: $\Gamma=(000)$, X$=(\frac{\pi}{a}00)$,
S$=(\frac{\pi}{a}\frac{\pi}{b}0)$, Y$=(0\frac{\pi}{b}0)$,
YZ$=(0\frac{\pi}{b}\frac{\pi}{c})$,
XYZ$=(\frac{\pi}{a}\frac{\pi}{b}\frac{\pi}{c})$, Z$=(00\frac{\pi}{c})$.}
\end{figure}

First, we consider magnetic couplings in the experimental crystal structures of
CuCl$_2$ and CuBr$_2$, as well as in the relaxed structure of chain-like
CuF$_2$. The DFT calculations of the band structure and the density of states
(DOS) of \cuxx\ compounds within the LDA yield a valence band width of
6--8\,eV,\cite{supp} in agreement with the experimental photoelectron
spectra.\cite{sawatzky1981} The valence band complex becomes slightly narrower
upon an increase in the ligand size, because the lower electronegativity of Cl
and Br brings the respective $p$ states closer to the Cu $3d$ states, thus
enhancing the hybridization and reducing the energy separation between the Cu
and ligand orbitals. All the band structures feature a separated band crossing
the Fermi level (Fig.~\ref{bands}). In the local-orbital representation
visualized by WFs (Fig.~\ref{J_analys}), this band is formed by the antibonding
$\sigma$*-combination of Cu $3d_{x^2-y^2}$  and X $p$ orbitals.\footnote{The
orbitals are denoted with respect to a local coordinate system, where for each
plaquette one of the Cu--X bonds and the direction perpendicular to the
plaquette are chosen as $x$ and $z$-axes, respectively.} The isolated
half-filled band suffices for describing the magnetic properties and the
low-lying magnetic excitations via the transfer integrals $t_i$ which are subsequently
introduced into a Hubbard model. Ligand valence $p$-orbital contributions to the 
magnetic orbital, denoted as $\beta$ in Table~\ref{T_tJ}, illustrate the increase in the
metal--ligand hybridization from F to Br. 

The dispersion calculated with the WF-based one-band TB model for CuBr$_2$ is also shown
in Fig.~\ref{bands}, and the leading transfer integrals together with the AFM
contributions $J_i^{\AFM}$ are given in Table~\ref{T_tJ}. The evaluation of
$J_i^{\AFM}$ requires the value of  $U_{\eff}$, which is not known precisely. Here,
we estimate $U_{\eff}$ by comparing the transfer integral $t_2$ obtained from
the TB analysis with the exchange coupling $J_2$ from the LSDA+$U$
calculations. While short-range couplings may involve large FM contributions,
the long-range coupling $J_2$ should be primarily AFM. Therefore,
$J_2^{\AFM}=J_2$ in a good approximation, and $U_{\eff}=4t_2^2/J_2$. This way,
we find $U_{\eff}=6$~eV for X = F, 4~eV for Cl, and 3~eV for Br. The reduction
in $U_{\eff}$ reflects the general trend of the enhanced Cu--X hybridization
and covalency, because the $U_{\eff}$ value pertains to the screened Coulomb
repulsion in the mixed Cu--X band. The enhanced hybridization leads to a
stronger screening, larger spatial extension and, thus, to the lower $U_{\eff}$ values.

\begin{table*}[tbp]
\begin{ruledtabular}
\caption{\label{T_tJ}\label{T_lsdau} Results for the experimental (X = Cl, Br)
and hypothetical (X = F) structures of CuX$_2$: the bridging angle $\theta$, the
ligand contribution to the magnetic orbital $\beta$, transfer integrals $t_i$, AFM
contributions to the exchange $J_i^{\AFM}=4t_i^2/U_{\eff}$, total exchange
integrals $J_i$ from LSDA+$U$ calculations with $U_d=7\pm 0.5$~eV, and the
effective on-site Coulomb repulsion $U_{\eff}$ obtained by equilibration of
$4t_2^2/U_{\text{eff}}$ (see text for details).
}
\begin{tabular}{c c c c c r r r r r}
& $\theta$ & $\beta$ & $t_1$ & $t_2$ & $J^{AFM}_{1}$ & $J^{AFM}_{2}$ & $J_1$ & $J_2$ & $U_{\text{eff}}$ \\
&  (deg)   &         & (meV) & (meV) & (meV)           & (meV)           & (meV)   & (meV)   &    (eV)          \\\hline
\cubr &  92   & 0.28 &  47 & 136 &  2.5 & 21.0 & $-8.8\pm 0.4$ & $22.2\pm 3.4$    & 3.0  \\ 
\cucl &  93.6 & 0.26 &  34 & 117 &  1.1 & 13.7 & $-12.9\pm 0.9$ & $13.4\pm 2.2$   & 4.0  \\
\cuf  & 102   & 0.15 & 132 &  50 & 11.6 &  1.6 & $5.4\pm 0.9$ & $1.2\pm 0.2$       & 6.0  \\
\end{tabular}
\end{ruledtabular}
\end{table*}

The estimates in Table~\ref{T_tJ} reveal two major differences between the
ionic CuF$_2$ and more covalent CuCl$_2$ and CuBr$_2$ compounds. First, the
nearest-neighbor (NN) coupling $J_1$ is AFM in the fluoride, while FM in the
chloride and bromide. Second, the AFM next-nearest-neighbor (NNN) coupling
$J_2$ is enhanced upon increasing the covalency of the Cu--X bonds. In CuF$_2$, this
coupling is weak ($J_2\ll J_1$), whereas in the chloride and bromide $J_2\geq
|J_1|$. The NNN coupling is amplified by the larger ligand size
and the increased covalency. This coupling involves the long-range Cu--X--X--Cu
pathway and requires a strong overlap between the ligand orbitals, which is
possible for X = Cl and especially Br, while remaining weak for the smaller
fluoride anion. The changes in the NN coupling seem to be well described by 
the GKA rules. Considering the trends for copper oxides,\cite{braden} one
expects FM $J_1$ for $\theta$ close to $90^{\circ}$, as in CuCl$_2$ and
CuBr$_2$, and AFM $J_1$ for $\theta>98^{\circ}$, as in the chain-like structure
of CuF$_2$. Nevertheless, the covalency is also paramount for the sign of
$J_1$, as shown by the magnetostructural correlations presented below
(Sec.~\ref{sec:bridging}).

Finally, we briefly compare our DFT-based estimates of $J_i$ with the
experiment. Because the chain-like polymorph of CuF$_2$ has not been prepared
experimentally, no comparison can be performed. The microscopic analysis of
CuCl$_2$ presented in Ref.~\onlinecite{FHC_CuCl2_DFT_chiT_simul_TMRG} shows
reasonable agreement between the experimental ($J_1=-7.8$~meV, $J_2=11.6$~meV) and
calculated ($J_1=-12.9\pm 0.9$~meV, $J_2=13.4\pm 2.2$~meV) values. The same is true 
for \cubr, where we evaluated the intrachain couplings as $J_1=-8.8\pm 0.4$~meV, 
$J_2=22.2\pm 3.4$~meV which compare well with recently published experimental data 
$J_1=-11.0\pm 1.6$~meV, $J_2=31.0$~meV.\cite{cubr2_2012} Moreover, our calculations 
reveal significantly lower deviations from experiment than those supplied in Ref.~\onlinecite{cubr2_2012}. 

Puzzled by the origin of the discrepancy between our values for $J_1$ and $J_2$ and the published 
calculational results for \cubr,\cite{cubr2_2012} we repeated the DFT+$U$ calculations for \cubr\ as well as 
\cucl\ with the code \textsc{vasp}\cite{vasp1,*vasp2} and the same computational parameters as used in 
Ref.~\onlinecite{cubr2_2012}. For the parameters $U_d$ and $J_d$, we adopted 8\,eV and 1\,eV, respectively, which corresponds to the effective $U\!=U_d\!-\!J_d\!=\!7$\,eV in Refs.~\onlinecite{cubr2_2012}. 
For the GGA+$U$ calculations, we used again a unit cell quadrupled along the $b$ 
axis and the $k$-mesh of 64 points. The resulting $J_1$ and 
$J_2$ values generally agree with the published values,\cite{banks2009,cubr2_2012} except for $J_1$ in CuBr$_2$, for 
which we obtain only half of the value provided in Ref.~\onlinecite{cubr2_2012}. The agreement with the 
experimental data can be improved by increasing the $U_d$ value. In particular, $U_d\!=\!12$\,eV 
yields $J_1\!=\!-95$\,K and $J_2$\,=\,113\,K for CuCl$_2$ and $J_1\!=\!-124$\,K and $J_2$\,=\,357\,K 
for CuBr$_2$, very close to the experimental estimates.\cite{FHC_CuCl2_DFT_chiT_simul_TMRG,cubr2_2012} This $U_d$ value is significantly higher than the $U_d=7$\,eV we used in our \textsc{fplo9.00-34} calculations.\footnote{A $U_d$ value of 7\,eV has turned out to supply good agreement with experimental data for several Cu$^{2+}$-compounds, see e.g. Refs.~\onlinecite{CaYCuO, dioptase, linaritePRB}.} There are basically two reasons for the large difference: The first reason are the different basis sets of \textsc{fplo9.00-34} and \textsc{vasp}, implementing local orbitals and projected augmented waves,\cite{paw1,*paw2} respectively, which crucially affect the local quantity $U_d$. Second, we used an around mean field double counting correction (DCC) while a fully localized limit DCC, which is always used in \textsc{vasp}, requires larger $U_d$ values.\cite{tsirlin2010_cucl}

\subsection{Variation of the bridging angle}
\label{sec:bridging}
\begin{figure}[tbp]
\includegraphics[width=8.6cm]{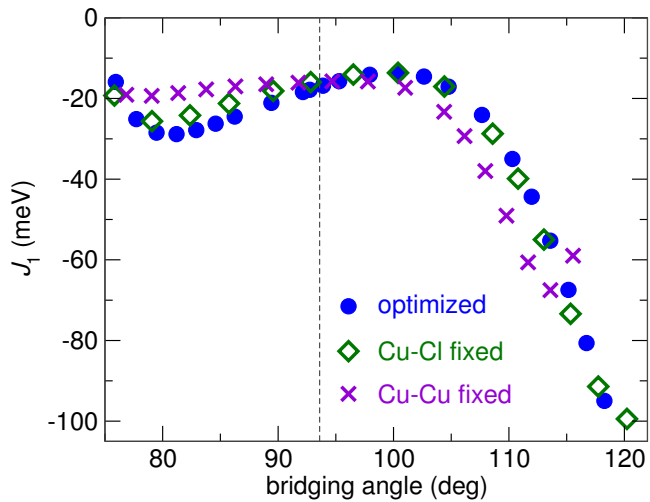}
\caption{\label{J_ang_o}(Color online) $J_1$ of \cucl\ as function of the
bridging angles where different structural parameters are fixed: i) the Cu-Cu distance is varied, while the X position is
subsequently optimized to yield the equilibrium Cu--X distance and $\theta$;
ii) the Cu-Cu distance is fixed, while the Cu--X distance is varied; and iii)
the Cu--X distance is fixed, while the Cu--Cu distance is varied. The
dashed vertical line indicates the experimental bridging angle.}
\end{figure}

To establish magnetostructural correlations in CuX$_2$ halides, we
systematically vary the bridging angle $\theta$ and evaluate the NN coupling
$J_1$. Since the Cu--Cu distance and two Cu--X distances form a triangle with
$\theta$ being one of its angles, the change in $\theta$ alters either the
Cu--Cu distance, or the Cu--X distance, or both. We compared different flavors
of varying $\theta$:\footnote{
we adopted a fictitious  structure, where the edge-sharing chains are simply
stacked, leading to a rectangular unit cell. This has the advantage of a much
simpler construction of the different structures.  For \cubr, also the small
tilting between the chains is not considered and the Cu--Br distance is
slightly enhanced to 2.45\,\r{A} enabling to span a broader range of the
bridging angles without getting artefacts from unphysically small Br--Br
distances.} i) the Cu-Cu distance is varied, while the X position is
subsequently optimized to yield the equilibrium Cu--X distance and $\theta$;
ii) the Cu-Cu distance is fixed, while the Cu--X distance is varied; and iii)
the Cu--X distance is fixed, while the Cu--Cu distance is varied. For all three
cases, we evaluated $J_1$ as a function of the Cu--X--Cu angle. Fig.~\ref{J_ang_o} 
shows on the example of \cucl\ that despite minor numerical differences, all three
methods conform well to each other. Additionally, we studied the influence of
$U_d$ by varying it in the wide range of 4--9\,eV. This causes a shift of the
curves along the vertical axis, but the qualitative behavior of $J_1$ versus
the Cu-X-Cu angle is retained.\cite{supp}

Remarkably, $J_1$ reaches its minimum absolute value at around
$\theta=100^{\circ}$ and becomes strongly FM at large bridging angles
(Fig.~\ref{J_ang_o}). This result is robust with respect to the particular
procedure of varying $\theta$. To better understand the microscopic origin of
this peculiar behavior, we performed similar calculations for CuF$_2$ and
CuBr$_2$. As different procedures of varying $\theta$ arrive at similar
results, we fixed the Cu--X distance for each ligand and achieved different
$\theta$ values by adjusting the Cu--Cu distance, only.  

\begin{figure*}[tbp]
\includegraphics[width=17.1cm]{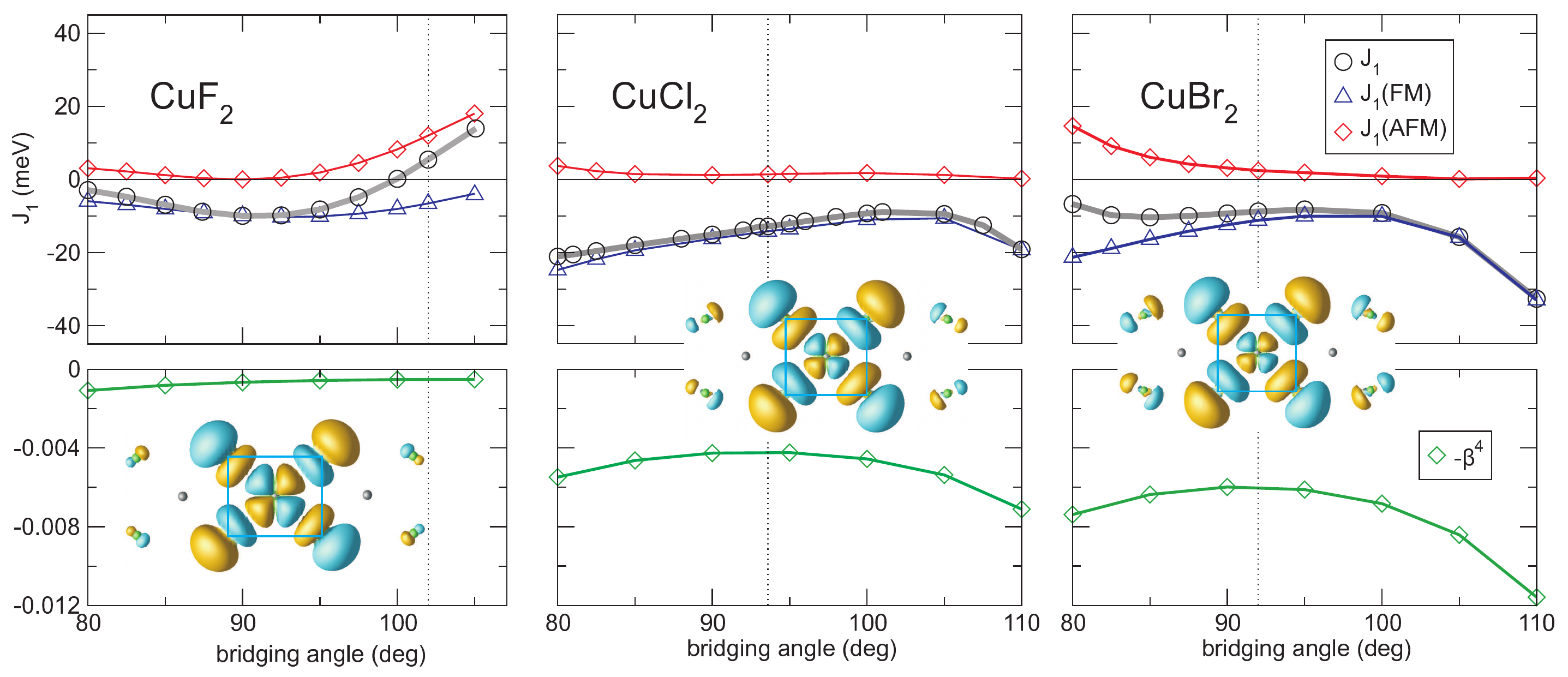}
\caption{\label{J_analys}(Color online) Magnetostructural correlations for the
CuX$_2$ halides (the Cu--X distance is fixed, the Cu--Cu distance is variable).
The upper panels show the total exchange $J_1$ (LSDA+$U$, $U_d=7$~eV) along
with $J_1^{\AFM}=4t_1^2/U_{\eff}$ and $J_1^{\FM}=J_1-J_1^{\AFM}$. The bottom
panels show $\beta^4$, where $\beta$ is the ligand's contribution to the
Cu-based magnetic orbital. The WFs for the experimental (relaxed) geometries are shown as
insets.}
\end{figure*}

Similar to our results for the fixed geometries (Table~\ref{T_tJ}),
magnetostructural correlations for $J_1$ (Fig.~\ref{J_analys}) reveal a large
difference between the ionic CuF$_2$ and covalent CuCl$_2$ and CuBr$_2$. In
CuF$_2$, $J_1$ follows the anticipated behavior with the FM-to-AFM crossover at
$\theta\simeq 100^{\circ}$. However, the covalent compounds always show FM
$J_1$, with a maximum (i.e., the minimum in the absolute value) at
$\theta\!=\!100\!-\!105^{\circ}$ and the enhanced FM character at even larger bridging angles.
This trend persists up to at least $\theta=120^{\circ}$ (Fig.~\ref{J_ang_o}).

The effect of strongly FM $J_1$ in CuCl$_2$ and CuBr$_2$ can be explained by
considering individual contributions to the exchange. The AFM contribution
$J_1^{\AFM}$ arises from the electron hopping between the Cu sites. The hopping
probability measured by the transfer integral $t_1$ critically depends on the
Cu--X--Cu bridging angle. In a simple ionic picture, the transfer is maximal at
$\theta=180^{\circ}$ (singly bridged) and approaches zero at $\theta=90^{\circ}$, thus providing the
microscopic reasoning behind the GKA rules. This anticipated trend is indeed
shown by CuF$_2$, where $J_1^{\AFM}=4t_1^2/U_{\eff}$ increases above
$\theta=90^{\circ}$ and underlies the increase in $J_1$. However, the covalent
CuX$_2$ halides show qualitatively different behavior with the very low (and
decreasing) $t_1$ and $J_1^{\AFM}$ up to at least $\theta=110^{\circ}$. This
result implies that the large contribution of the ligand states in a covalent
compound has also a strong influence on the Cu--X--Cu hopping process and alters the
anticipated trend for the AFM exchange.

The FM contribution $J_1^{\FM}$ can be evaluated as \mbox{$J_1-J_1^{\AFM}$}, where we
use $J_1$ from the LSDA+$U$ calculation and $J_1^{\AFM}=4t_1^2/U_{\eff}$ from
the TB analysis. Microscopically, $J_1^{\FM}$ originates from the Hund's
coupling on the ligand site\cite{beta4} and/or from the FM coupling between the
Cu $3d$ and ligand $p$ states.\cite{FHC_LiCuVO4_MH_DFT_DMRG,*kuzian2012}
Regarding the former mechanism,\cite{beta4} a simple model expression reads as
$J_1^{\FM}=-\beta^4J_H$, where $\beta$ is the ligand's contribution to the
Cu-centered magnetic orbital, and $J_H$ is the (effective) Hund's coupling on the ligand. Even
though this expression is derived for $\theta=90^{\circ}$, our data obtained
for different $\theta$ values are well understood in terms of the variable
$\beta$ (see bottom panels of Fig.~\ref{J_analys}). The increase in the
bridging angle leads to larger $\beta$, thus enhancing $J_1^{\FM}$. Since
$\beta$ enters $J_1^{\FM}$ as $\beta^4$, its effect should be dominant over any
other contributions, such as slight variations of $J_H$. The increase in $\beta$ also explains the
increasing FM contribution at low $\theta$ (Fig.~\ref{J_analys}).

In contrast to the covalent chloride and bromide, the ionic CuF$_2$ shows only
a minor FM contribution owing to the very low $\beta$. We also tried to
artificially enhance $\beta$ by reducing the Cu--F bonding distance down to
1.60\,\r{A}. For bridging angles larger than 100$^{\circ}$ the AFM coupling becomes 
twice as large as for the Cu--F distance of 1.91\,\r{A} and for angles smaller than 
80$^{\circ}$ the model compound becomes also AFM. The FM coupling strength about 
90$^{\circ}$ is almost unaffected. This indicates the robust ionic
nature of Cu--F bonds. The reduction in the Cu--F distance increases the
electron transfer without changing the hybridization, hence $J_1^{\AFM}$ is
increased, while $J_1^{\FM}$ remains weak.

\subsection{Cluster models}
In a periodic calculation, the variation of structural parameters, such as bond
lengths and angles, is generally challenging: the high symmetry couples the
structural parameters to each other. As a result, changing a single parameter
is often impossible without affecting the other parameters. The cluster models
are more flexible and may allow for an independent variation of individual bond
lengths and angles. This property renders the clusters as an excellent
playground to study the magnetostructural correlations. 

\begin{figure}[tbp]
\includegraphics[width=8.6cm]{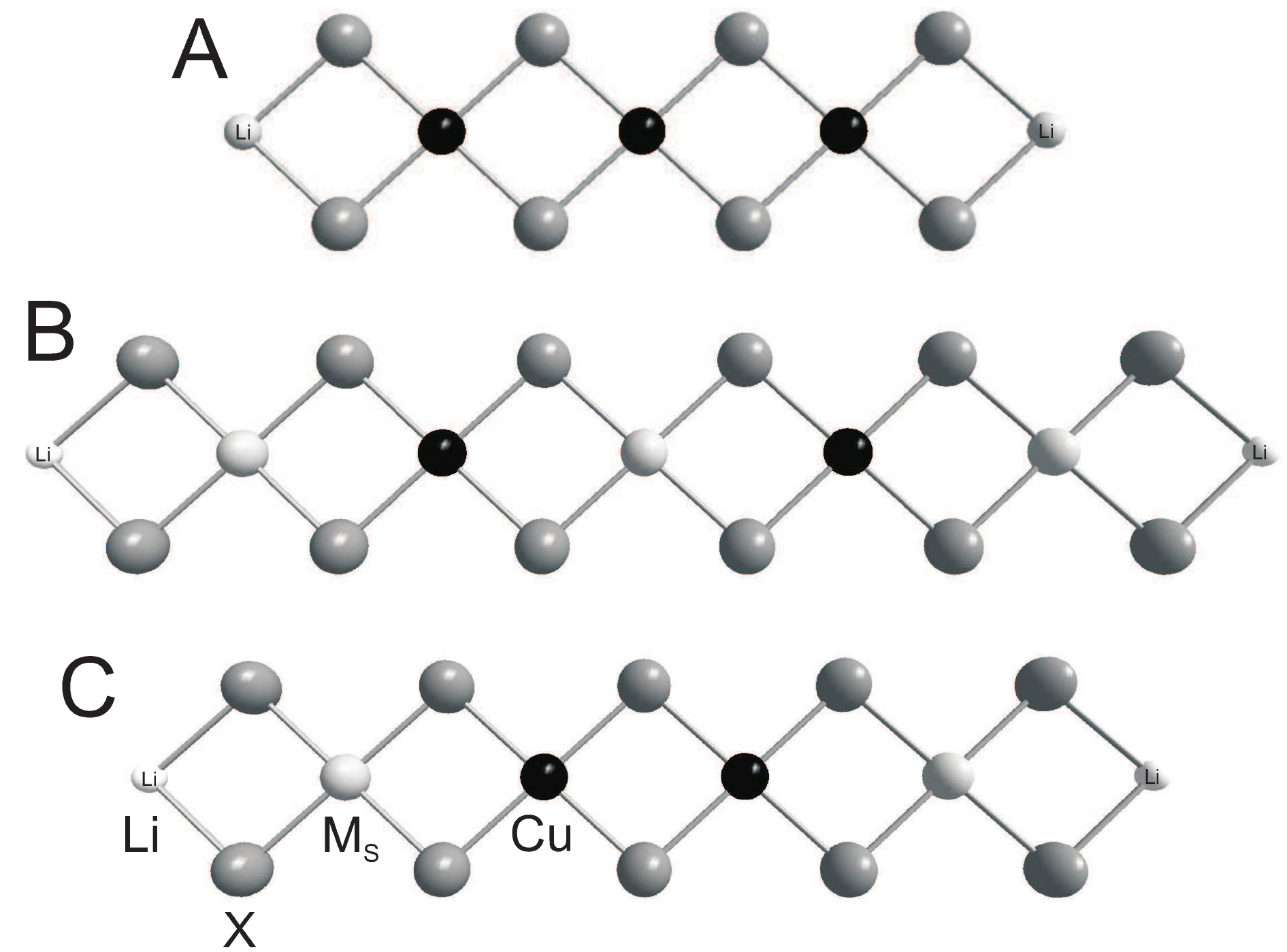}
\caption{\label{clusters} Three examples of model clusters: Cu$_3$ trimer
cluster as the minimal cluster for the evaluation of $J_1$ and $J_2$ (A); the
pentamer cluster for calculating $J_2$, with only two Cu$^{2+}$ and three
substituted non-magnetic ions (B); and the tetramer cluster for calculating
$J_1$ with two magnetic Cu and two nonmagnetic $M_S$ centers (C).}
\end{figure}

Before discussing the intrachain couplings  using a combination of periodic and
cluster models, we first want to demonstrate how cluster models for the three
Cu dihalide compounds are constructed. Since the chains are spatially
well-separated from each other, we can consider segments of a chain, with the
terminal Li atoms keeping the electroneutrality (Fig.~\ref{clusters}). No
additional point charges are required, so that the clusters are kept as simple
as possible.

\begin{figure}[tbp]
\includegraphics[width=8.6cm]{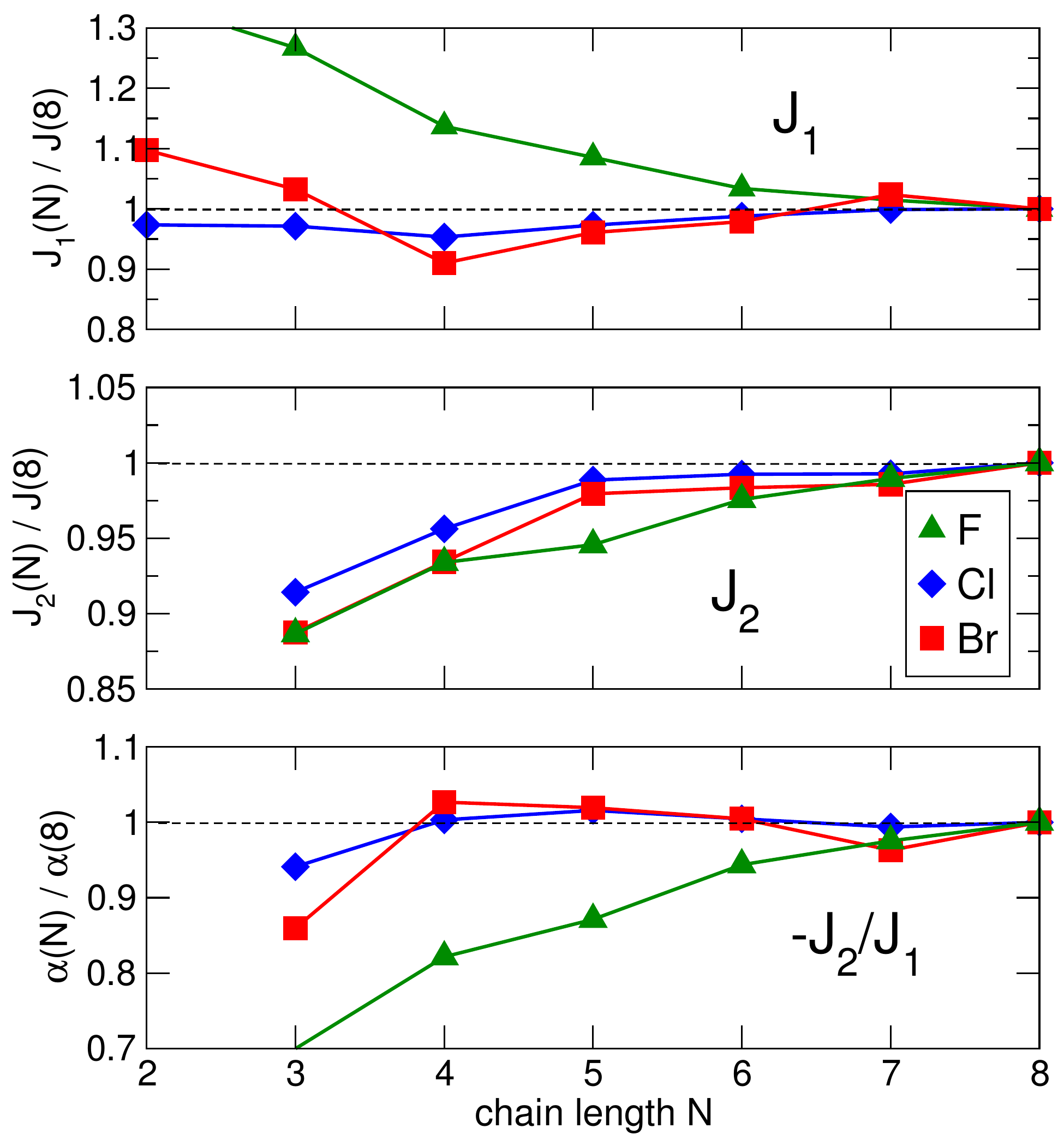}
\caption{\label{F-jN} (Color online) $J_1(N)/J_1(N\!=\!8)$, $J_2(N)/J_2(N\!=\!8)$ and
$\alpha(N)/\alpha(N\!=\!8)$ as a function of the chain length $N$. The
bridging angle is fixed to the experimental (CuCl$_2$ and CuBr$_2$) and
optimized (CuF$_2$) values, respectively. For $J_2$ and $-J_2/J_1$, the
minimal number of Cu-centers amounts to three. The exchange couplings are 
calculated with the LSDA+$U$ method with $U_d$\,=\,7\,eV.} 
\end{figure}

First, the effect of the chain length on $J_1$, $J_2$ and the ratio $-J_2/J_1$
is investigated (Fig.~\ref{F-jN}).  For all three compounds, small clusters,
such as dimers or trimers, are insufficient for describing the magnetic
properties.  The convergence with respect to the cluster size is different for
different compounds (e.g., the ionic \cuf\ demonstrates the slowest
size convergence).  To ensure a meaningful comparison with the periodic model
or the experimental data, the convergence with respect to the cluster size has
to be carefully checked. 

\begin{figure}[tbp]
\includegraphics[width=8.6cm]{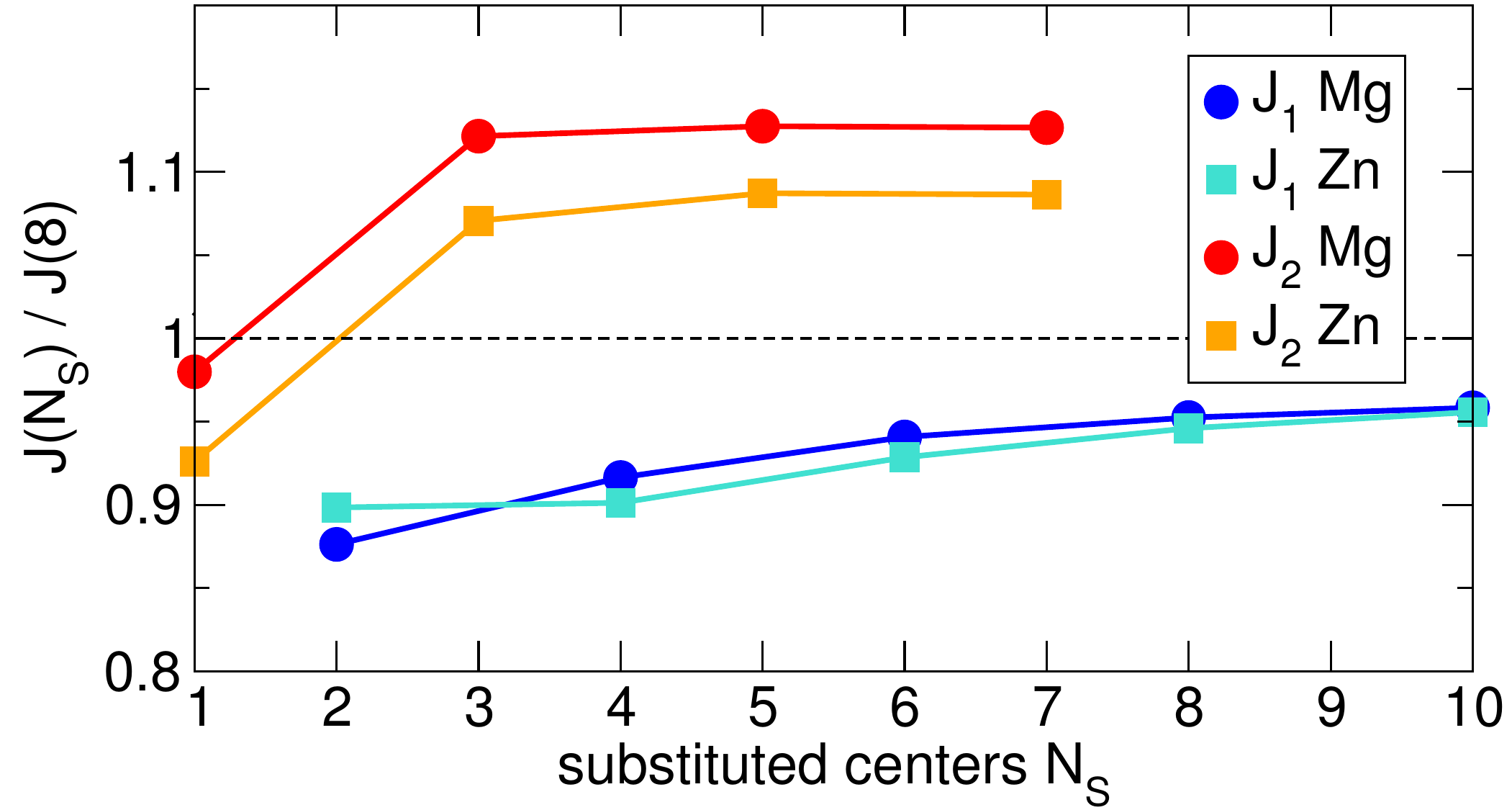}
\caption{\label{F-jN_sub}(Color online) $J_1$ and $J_2$ of the Cu-Br
clusters calculated with clusters containing two correlated and N$_S$
uncorrelated Mg$^{2+}$ or Zn$^{2+}$ centers. The bridging angle is fixed to the 
experimental value. The resulting exchange integrals are normalized to that for the
Cu$_8$-octamer cluster. For the calculations, the LSDA+$U$ method is used with $U_d$\,=\,7\,eV.} 
\end{figure}

On the other hand, a large number of correlated centers requires a large number
of spin configurations to estimate exchange couplings. While larger clusters
are still feasible for DFT, they may pose a problem for advanced \textit{ab
initio} quantum-chemical methods. Therefore, we attempted to reduce the number
of correlated Cu$^{2+}$ ions by substituting them by formally nonmagnetic
Mg$^{2+}$ and Zn$^{2+}$ ions (Fig.~\ref{clusters}). Even with this minimum
number of correlated centers, deviations below 10\% to the size-converged
Cu$_8$ octamer cluster are obtained for the Cu--Br (Fig.~\ref{F-jN_sub}) and
also for the Cu--Cl clusters. In case of Cu-F, where convergence is reached at larger cluster size, at least
four correlated centers are required to reduce the deviations down to that level.

Similar results, as for the $J$'s, concerning size convergence and
substitutions are obtained for the NN and NNN transfers, $t_1$ and $t_2$,
calculated in LDA. These results show that the simple clusters suffice for
describing the intrachain physics of these compounds and that the problem of
appropriately embedding the clusters may be at least partially bypassed by
increasing the cluster size and substituting part of the correlated centers
with weakly correlated ions.

\subsection{Cluster versus periodic models}
In the following, both cluster and periodic models will be used for calculating
$J_2$ and the $-J_2/J_1$ ratio, as well as the transfer integrals $t_i$ of the
Cu dihalides.  The comparison of periodic and cluster models for a broad range
of bridging angles allows to exclude an accidental agreement between both
models, which can be realized in a specific geometry by appropriately choosing
the chain length, substitutions, and the termination of the cluster. However,
when the cluster is prepared in such a way, the good agreement with the
periodic model would be lost by varying the geometrical parameters.

\begin{figure}[tbp]
\includegraphics[width=8.6cm]{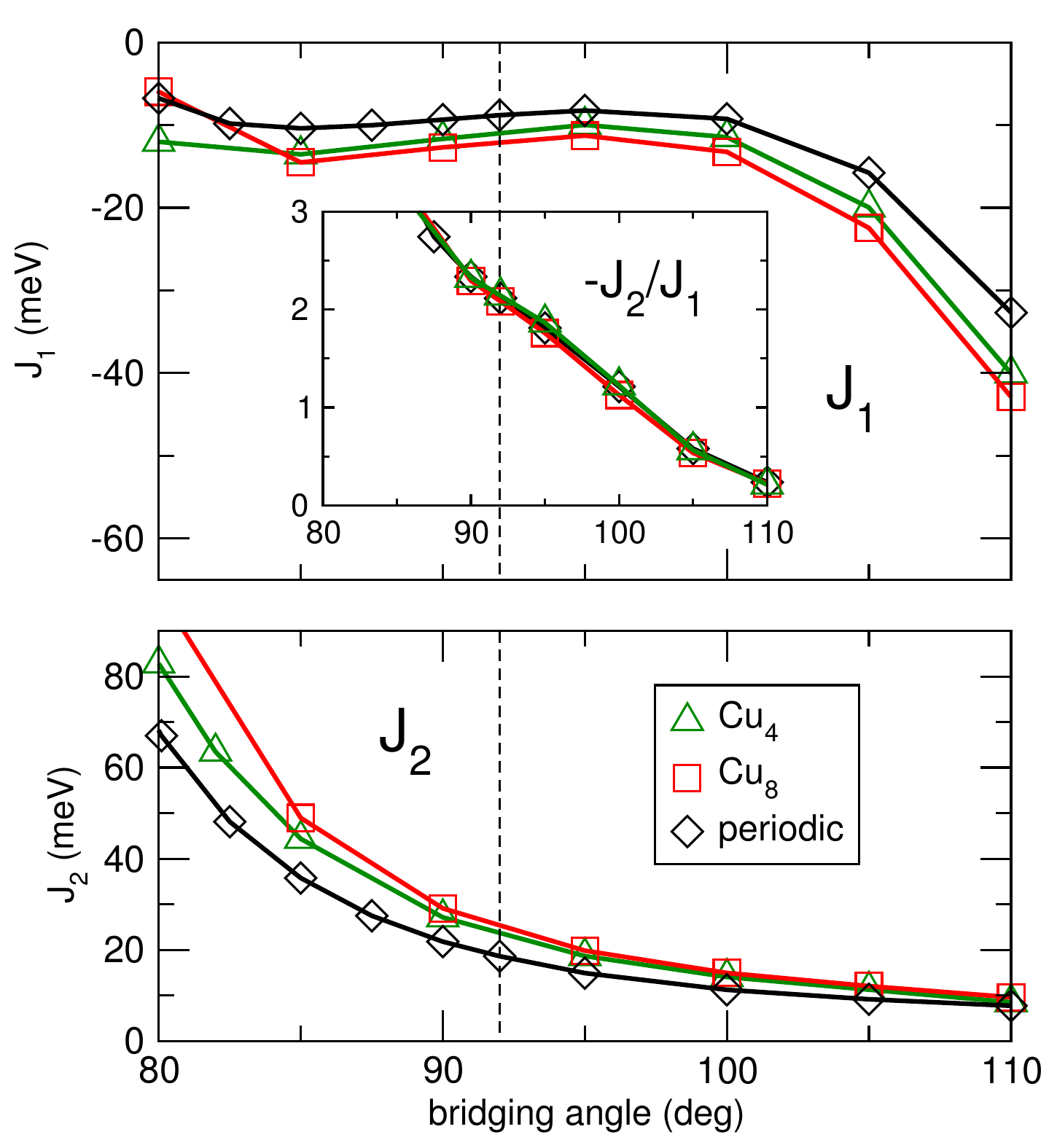}
\caption{\label{F-Jvar}(Color online) CuBr$_2$: exchange integrals $J_1$ and
$J_2$ as a function of the bridging angle. A periodic 
as well as two different cluster models (Cu$_4$ and Cu$_8$) were used. The
inset shows the ratio $-J_2/J_1$. The dashed vertical line indicates the
experimental bridging angle of 92$^{\circ}$. For the calculations, the LSDA+$U$ method is used with $U_d$\,=\,7\,eV.}
\end{figure}

\begin{figure}[tbp]
\includegraphics[width=8.6cm]{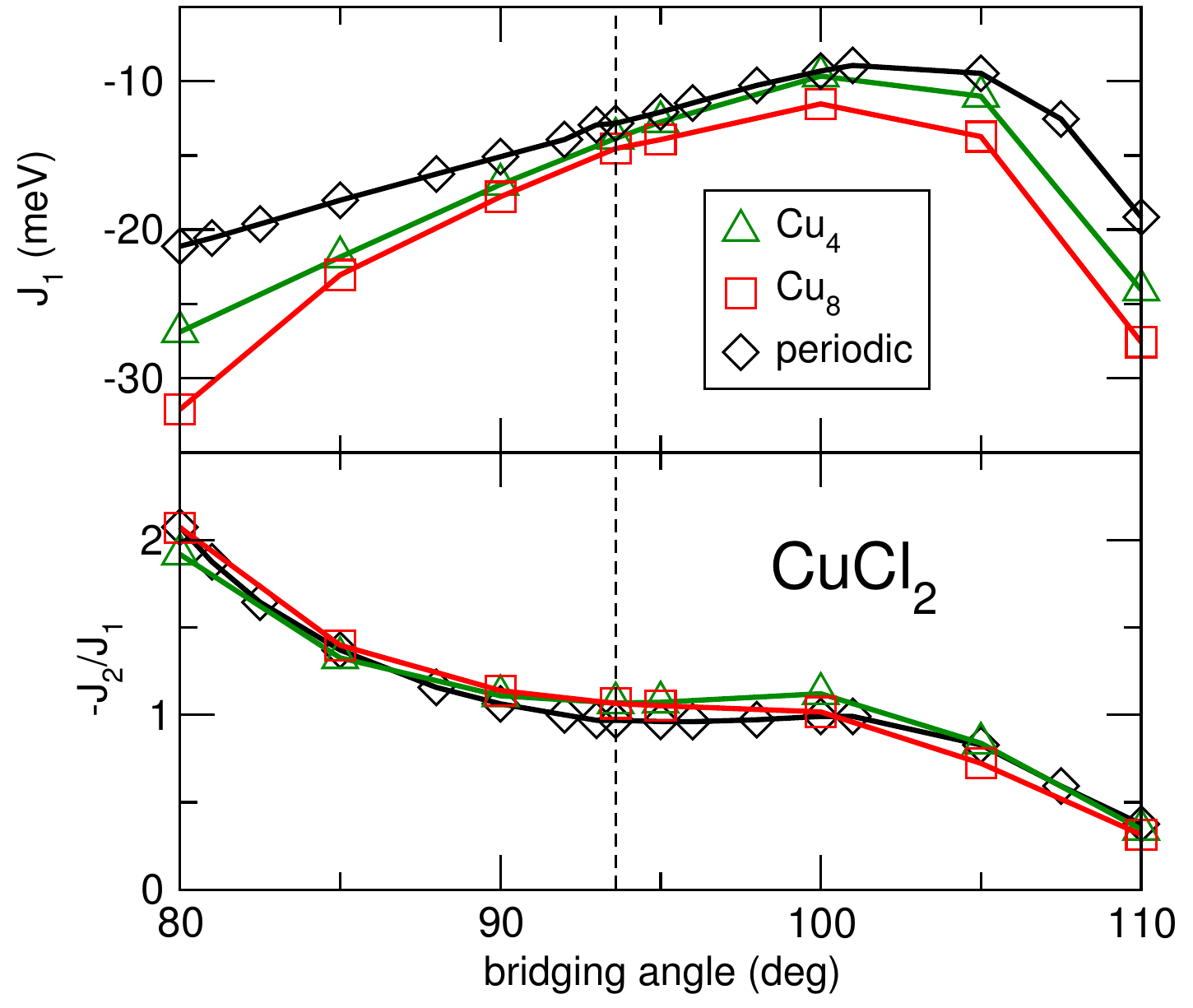}
\caption{\label{F-varcl}(Color online) Exchange integrals $J_1$
and the ratio $-J_2/J_1$ of \cucl\ as a function of the bridging angle
calculated with a periodic and two cluster models. The dashed vertical line
indicates the experimental bridging angle of 93.6$^{\circ}$. For the \cuf\
data, see supplementary information. For the calculations, the LSDA+$U$ method is used with $U_d$\,=\,7\,eV.}
\end{figure}

\begin{figure}[tbp]
\includegraphics[width=8.6cm]{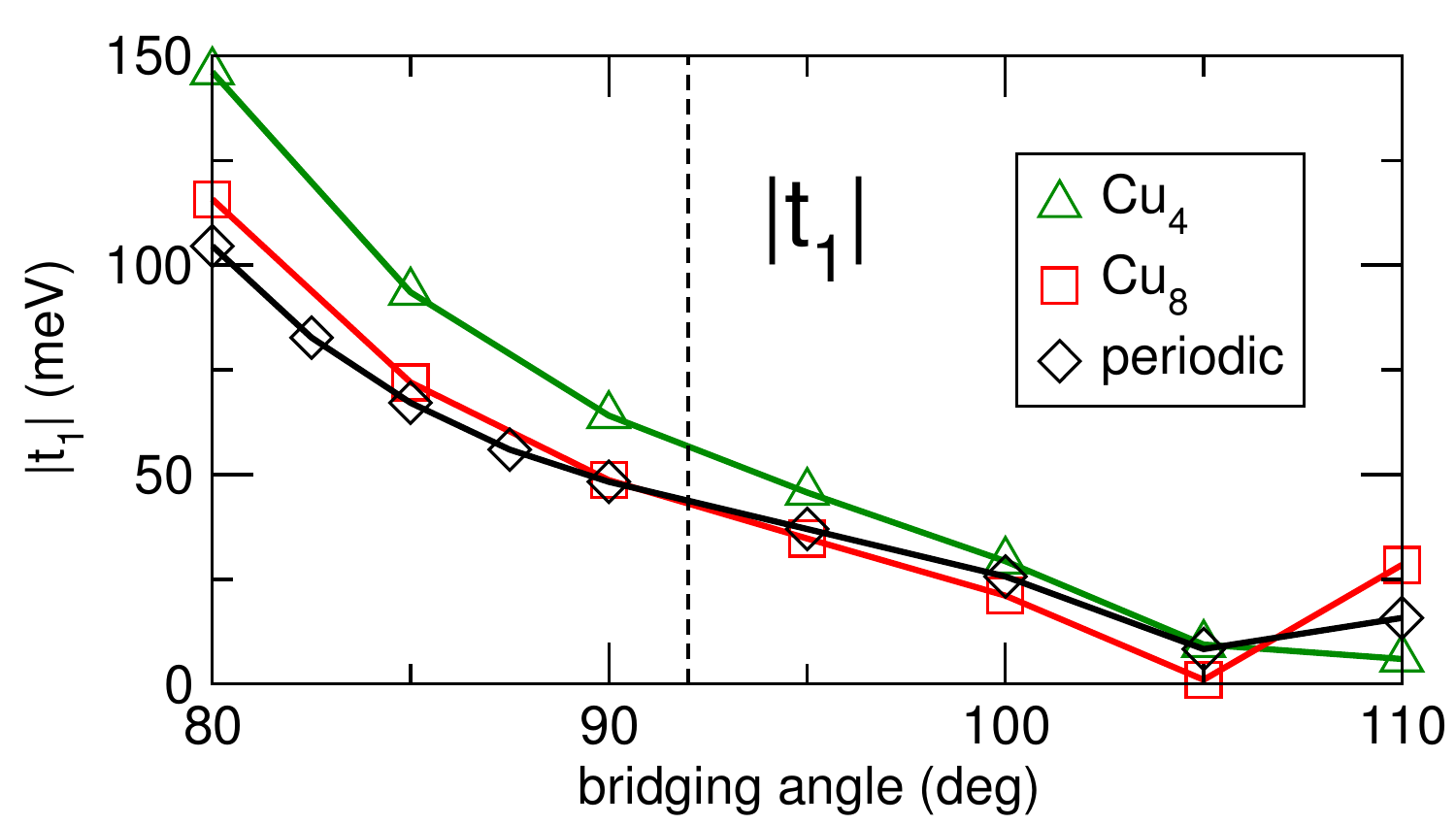}
\caption{\label{F-t1}(Color online) \cubr: the nearest-neighbor transfer
integral $t_1$ as a function of the bridging angle calculated with a periodic as well as two cluster models (Cu$_4$ and Cu$_8$).}
\end{figure}

The exchange integrals as well as the $-J_2/J_1$ ratio versus the bridging
angles are depicted in Figs.~\ref{F-Jvar} and~\ref{F-varcl} for CuBr$_2$
and CuCl$_2$, respectively. A comparison of the nearest-neighbor transfer
integral $t_1$ of \cubr, calculated with cluster and periodic models, is
shown in Fig.~\ref{F-t1}. The clusters can reproduce the results of band
structure calculations over the whole range of bridging angles, thus
justifying the construction of the clusters. In the $-J_2/J_1$ ratio,
which governs the magnetic ground state, the deviations between the
cluster and periodic models are compensated to a large degree. 

In \cuf, the deviations between $J_1$ and $J_2$ obtained in the cluster and
periodic models, respectively, are also compensated in the ratio $-J_2/J_1$,
except for the smallest bridging angles.\cite{supp} The singularity in
$-J_2/J_1$ at about 100$^{\circ}$ arises from the crossover between the FM and
the AFM $J_1$. 

These results show that well-controlled cluster models are capable of
describing local properties of ionic as well as strongly covalent solids,
whereas the good agreement with band structure calculations is not accidental
or artificial. Finally, the results demonstrate that superexchange and
magnetic coupling in insulators are relatively short-range effects even for
strongly covalent compounds.

\subsection{LSDA+$U$ vs. hybrid functionals}
\label{sec:luvsb3}
A common problem of DFT-based approaches applied to strongly correlated
electrons is the ambiguous choice of empirical parameters and corrections that are required to mimic
many-body effects, e.g., in the mean-field DFT+$U$ approach. Hybrid functionals
represent an alternative, although still empirical, way of simulating the effect 
of strong electron correlations within DFT. In this way, the non-local exact exchange is mixed
with the local LDA or GGA exchange, while the mixing parameter $\alpha$ is
typically the only free parameter.
In contrast to DFT+$U$, hybrid functionals
are more robust with respect to the adjustable parameters, and the constant
value of $\alpha=0.20$ or $\alpha=0.25$ can be used in a rather general fashion. Additionally, the 
exact exchange correction is generally applied to all orbitals while in DFT+$U$ 
the corrections are applied to a certain set of orbitals which are assumed to be 
the strongly correlated ones.

\begin{figure*}[tbp]
\includegraphics[width=17.1cm]{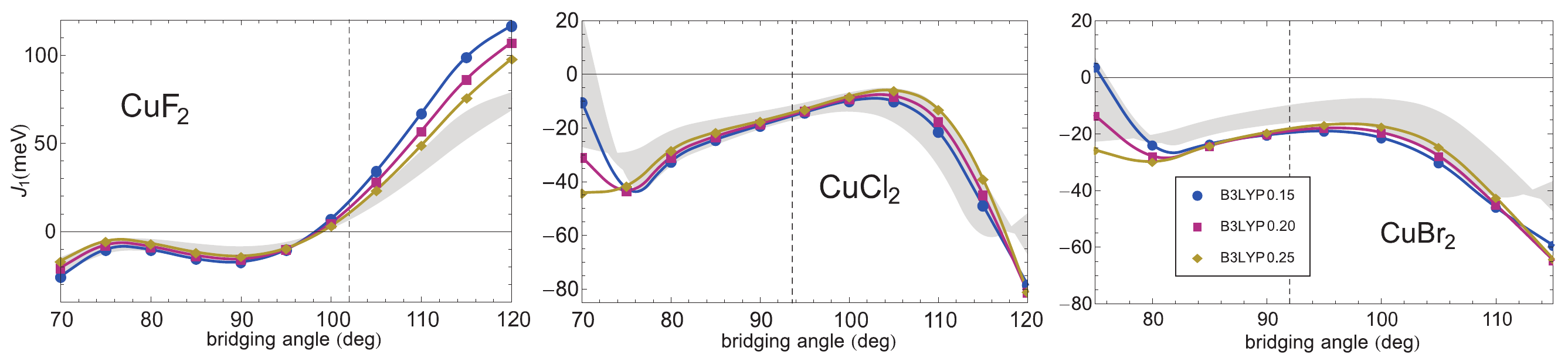}
\caption{\label{F-b3lyp}(Color online) The exchange integral $J_1$ of 
the \cuxx\ compounds as a function of the bridging angle. The calculations are
done for a dimer model with LSDA+$U$ and $U_d=7\pm 1$~eV (grey area),
and with the B3LYP functional ($\alpha=0.15-0.25$).}
\end{figure*}

In this study, we apply the B3LYP functional 
on dimer models and vary $\alpha$ between 0.15
and 0.25 ($\alpha=0.20$ corresponds to the standard B3LYP functional as
implemented in \textsc{gaussian}).
Although we pointed out that dimer models are too small for calculating $J_1$ in quantitative agreement with the periodic model, 
they are well suited for comparing the different DFT methods and parameter sets.\footnote{For larger clusters of 
\cubr\ the expectation values \mbox{$\langle S^2 \rangle$} of the broken symmetry (BS) states (which are described by single Slater determinants) calculated 
with the hybrid functional tend to deviate from the theoretical values. The deviations 
($<10$\%), depending on the bridging angle and $\alpha$, slightly shift the BS states and thus 
affect the exchange couplings. This impedes a fair comparison of the different methods and different choices of 
parameters what is exactly our goal.}  
Despite substantially different treatment of many-body effects in DFT+$U$ and hybrid functionals, 
the resulting exchange integrals of all three CuX$_2$ compounds are quite similar
(Fig.~\ref{F-b3lyp}). Thus, the B3LYP calculations confirm the LSDA+$U$ results, justify the 
choice of the free parameters in the latter approach and demonstrate that the unusual 
FM $J_1$ coupling of \cucl\ and \cubr\ is not an artifact of a certain method. Despite
the fact that B3LYP was originally constructed to reproduce the thermodynamical
data for small molecules, it provides meaningful results for strongly
correlated systems such as CuX$_2$, in line with the earlier
studies.\cite{ruiz97,munoz2002,b3lyp_QC1} Moreover, the calculated exchange
integrals are robust with respect to $\alpha$: the exchange integrals are
rather insensitive to the choice of this parameter.

\section{Discussion and Summary}
\label{summ}
Our study of magnetostructural correlations in the CuX$_2$ halides reveals the
crucial role of the ligand in magnetic exchange. Its effect is two-fold: First,
the larger size of Cl and Br is responsible for the enhanced NNN coupling $J_2$
that is assisted by the sizable overlap of ligand $p$ orbitals along the
Cu--X--X--Cu pathway. Second, the covalent nature of the Cu--Cl and Cu--Br
bonds underlies the large ligand contribution to the magnetic orbitals and,
consequently, the strong FM nearest-neighbor (NN) coupling $J_1$ in the broad range of bridging
angles which could be ascribed to Hund's exchange on the ligand site. The tendency 
of covalent Cu$^{2+}$ halides to exhibit FM exchange along the
Cu--X--Cu pathways can be illustrated but also challenged by several experimental
observations. It should be emphasized that ferromagnetic NN coupling requires not 
only sizeable ferromagnetic contributions but also small transfer integrals as were 
found for \cucl\ and \cubr. Otherwise, the AFM contributions will outweigh the FM 
terms even for covalent compounds.

Experimental data for Cu$^{2+}$ chlorides and bromides indeed show the robust FM NN coupling for
the bridging angles below $90^{\circ}$. While the $\theta<90^{\circ}$ regime is
not typical for the ionic oxides and fluorides, it is abundant in covalent
systems and observed, e.g., in Cu-based FM spin
chains.\cite{FM_wl,*devries1987} The FM nature of the NN coupling at
$\theta\!=\!90\!-\!95^{\circ}$ is evidenced by CuCl$_2$ and CuBr$_2$
themselves.\cite{banks2009,FHC_CuCl2_DFT_chiT_simul_TMRG,zhao2012,cubr2_2012} However,
larger $\theta$ values are less common and require geometries other than the
edge-sharing CuX$_2$ chains considered in the present study. 

The angles of $\theta>95^{\circ}$ are only found in edge-sharing dimers and
corner-sharing chains. Moreover, the respective experimental situation is
rather incoherent. In (CuBr)LaNb$_2$O$_7$ and (CuCl)LaTa$_2$O$_7$, the
corner-sharing geometry with $\theta>100^{\circ}$ indeed leads to the FM
exchange, although with a tendency towards AFM exchange at $\theta\geq
108-109^{\circ}$ (Refs.~\onlinecite{cubr-nb,cucl-ta}). By contrast, the
Cu$_2$Cl$_6$ dimers may reveal the AFM exchange even at $\theta\simeq
95.5^{\circ}$, as in LiCuCl$_3\cdot 2$H$_2$O (Ref.~\onlinecite{abrahams1963}) 
or TlCuCl$_3$ and KCuCl$_3$ (Ref.~\onlinecite{shiramura1997}) where the latter 
exhibit transfer integrals that are 3.5 times larger as that in \cucl. On the 
other hand, similar Cu$_2$Cl$_6$ dimers with the same bridging angle of
$\theta\simeq 95.5^{\circ}$ in the spin-ladder compound IPA-CuCl$_3$ feature the
sizable FM intradimer coupling.\cite{masuda2006}

These experimental examples show that the bridging angle $\theta$
may not be the single geometrical parameter determining the Cu--X--Cu
superexchange. Details of the atomic arrangement are important even for
Cu$^{2+}$ oxides,\cite{ruiz97,*ruiz97_2} whereas in more covalent systems
this effect is likely exaggerated because interactions involve specific
orbitals, so that each bond determines the orientation of other bonds around
the same atom. 
We have pointed out, that such magnetostructural correlations, essential for understanding the 
magnetic behavior and for the search of new interesting materials, can nicely be 
investigated with cluster models. In particular in case of intricate crystal structures clusters 
enable studying effects of each structural parameters separately, while for periodic models 
only a set of parameters can be modified at once.

On a more general side, our results identify the Cu--X--Cu pathways as the
leading mechanism of the short-range exchange in Cu$^{2+}$ halides. The fact
that the magnetostructural correlations weakly depend on the procedure of
varying $\theta$ (Fig.~\ref{J_ang_o}) entails the minor role of direct Cu--Cu
interactions, because the coupling always evolves in a similar fashion, no
matter whether the Cu--Cu distance is fixed or varied. Therefore, the nature of
the ligand is of crucial importance, and affects the Cu--X--Cu hopping along with the
FM contribution, presumably related to the Hund's coupling on the ligand
site.\cite{beta4} In ionic systems, the nearest neighbor 
hopping increases with the bridging angle and dominates over the small FM contributions, thus
leading to the conventional GKA behavior. However, the GKA behavior may be strongly altered in covalent
compounds, as shown by our study and previously argued in model studies on the effect of side
groups and distortions.\cite{gk,leb2,ruiz97}


From the computational perspective,
magnetic modeling of chlorides and bromides is generally challenging.
Although these compounds are still deep in the insulating regime, far from the
Mott transition ($t_i\ll U_{\eff}$, see Table~\ref{T_tJ}), the
sizable hybridization of ligand states with correlated Cu $3d$ orbitals
challenges the DFT+$U$ approach, with correlation effects restricted to the $d$ states. The microscopic
evaluation of magnetic couplings in Cu$^{2+}$ chlorides and bromides indeed
leads to large uncertainties.\cite{cucl-ta,Cs2CuX4_DFT} Hybrid functionals, on the other hand, tend to overestimate magnetic exchange couplings\cite{munoz2002} and provide a working, but empirical solution to the problem of strongly correlated electronic systems. This calls for the development and application of alternative 
techniques, as for instance \textit{ab
initio} quantum-chemical calculations, appropriately accounting for strong 
electron correlations. Since the wavefunction-based quantum-chemical calculations are
presently restricted to finite systems, they require the construction of
appropriate clusters. This task has been successfully accomplished in our work.
We have demonstrated that relatively small clusters with a low number of
correlated centers are capable of reproducing the results obtained for periodic
systems, and provide adequate estimates of the magnetic exchange even for the 
long-range Cu--X--X--Cu interactions.

In summary, we have studied magnetostructural correlations in the family of
CuX$_2$ halides with X = F, Cl, and Br. Our results show substantial
differences between the ionic CuF$_2$ and largely covalent CuCl$_2$ and
CuBr$_2$. The fluoride compound behaves similar to Cu$^{2+}$ oxides, and shows
weak FM exchange at the bridging angles close to $\theta=90^{\circ}$ along
with the AFM exchange at $\theta\geq 100^{\circ}$. Going from F to Cl and Br 
leads to two major changes: i) the larger size of the ligand
amplifies the AFM next-nearest-neighbor coupling $J_2$; ii) the increased
covalency of the Cu--X bonds results in the strong mixing between the Cu $3d$
and ligand $p$ states, and enhances the FM contribution to the short-range nearest-neighbor
coupling $J_1$. 
We have constructed cluster models which, first, supplied an excellent description 
of local properties of the solids. Second, they turned out as highly valuable 
tool for investigating magnetostructural correlations, e.g., they could be instrumental 
in the microscopic analysis of the covalent Cu$^{2+}$ chlorides and bromides with 
interesting but still barely explored magnetism.
Finally, they seem to be a viable approach to parameter-free quantum-chemical 
calculations of strongly correlated solids.

\section{Acknowledgements}
We acknowledge valuable discussions with O. K. Anderson and P. Blaha.  S. L.
acknowledges the funding from the Austrian Fonds zur F\"orderung
der wissenschaftlichen Forschung (FWF). A.T. was partly supported by the
Mobilitas grant of the ESF.

\bibliographystyle{apsrev4-1}
%

\clearpage


\begin{table*}[h]
\begin{tabular}{c}
\huge{\texttt{Supporting Material}} \\
\end{tabular}
\end{table*}

\begin{figure*} [h]
\includegraphics[width=14cm]{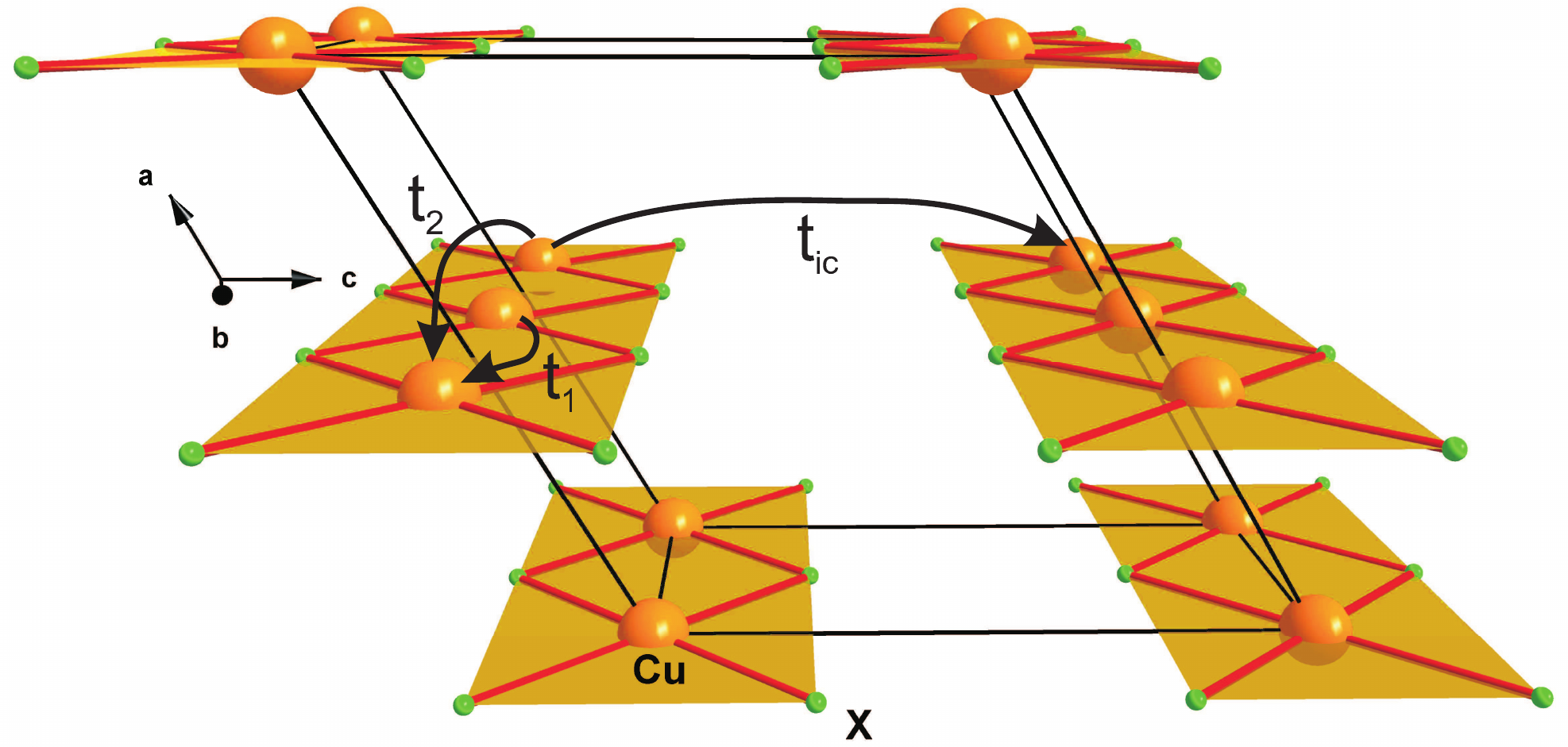}
\begin{flushleft}
(Color online) Crystal structure of the \cuxx\ compounds. Edge-sharing CuX$_4$ plaquettes form planar chains running along the b-axis. For \cuf, this is a fictitious structure which is introduced in order to investigate the effects of different ligand size on the intrachain magnetic couplings. Real \cuf\ features a 2-dimensional structure of corner-sharing CuF$_4$-plaquettes that would be inappropriate for such purposes. $t_1$ and $t_2$ denote nearest-neighbor and next-nearest-neighbor intrachain hopping, respectively. $t_ic$ is the interchain hopping which is expected to have a negligible effect on the intrachain couplings.
\end{flushleft}
\end{figure*}

\begin{figure*}[tbp]
\includegraphics[width=9cm]{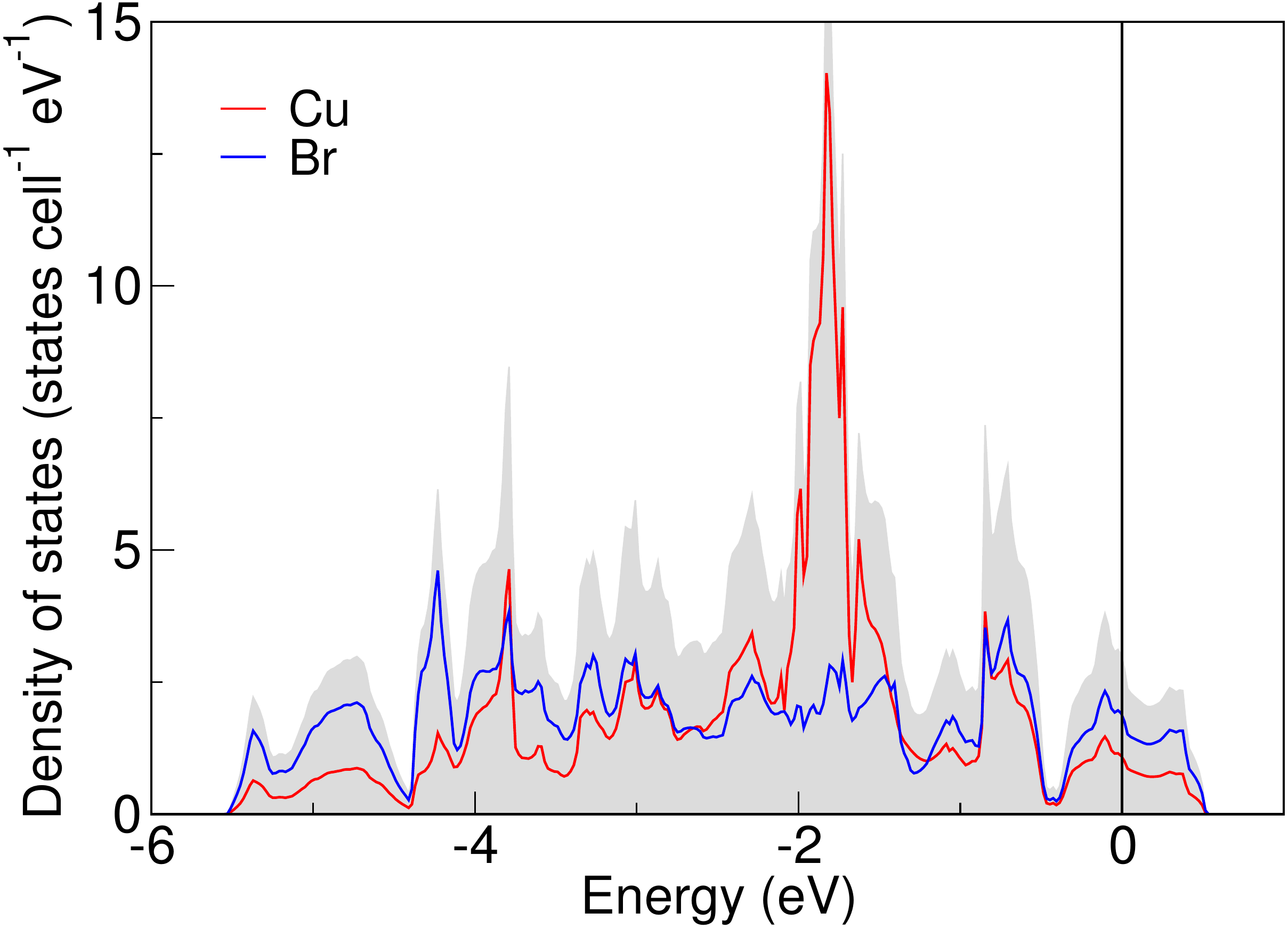}
\begin{flushleft}
(Color online) Density of states (DOS) around the Fermi level of \cubr, calculated with LDA. The gray shaded area corresponds to the total DOS and the red and blue lines belong to the partial Cu(3$d$) and Br(4$p$) DOS, respectively.
\end{flushleft}
\end{figure*}

\begin{figure*}[tbp]
\includegraphics[width=9cm]{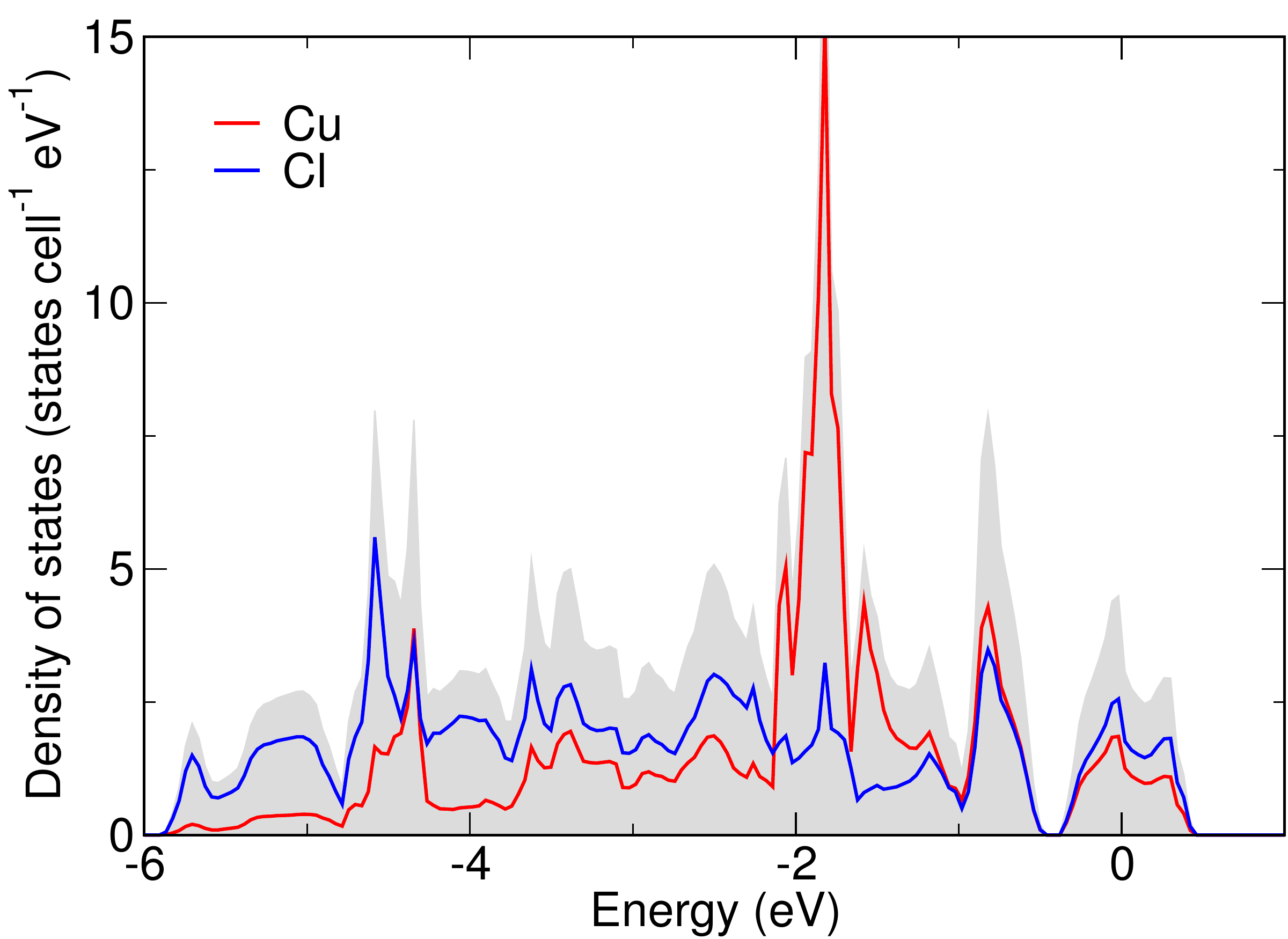}
\begin{flushleft}
(Color online) Density of states (DOS) around the Fermi level of \cucl, calculated with LDA. The gray shaded area corresponds to the total DOS and the red and blue lines belong to the partial Cu(3$d$) and Cl(3$p$) DOS, respectively.
\end{flushleft}
\end{figure*}

\begin{figure*}[tbp]
\includegraphics[width=9cm]{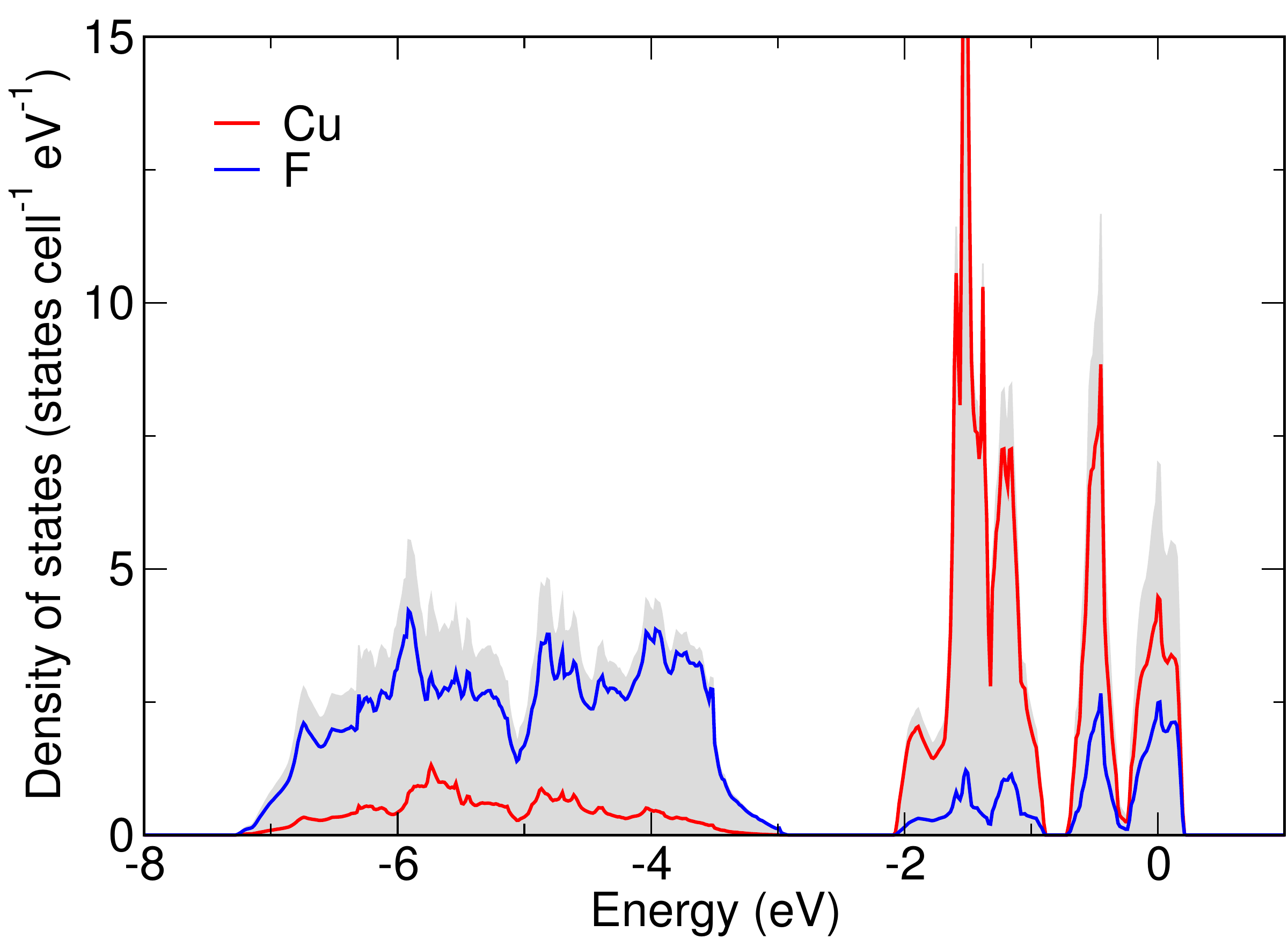}
\begin{flushleft}
(Color online) Density of states (DOS) around the Fermi level of \cuf, calculated with LDA. The gray shaded area corresponds to the total DOS and the red and blue lines belong to the partial Cu(3$d$) and F(2$p$) DOS, respectively.
\end{flushleft}
\end{figure*}

\begin{figure*}[tbp]
\includegraphics[width=9cm]{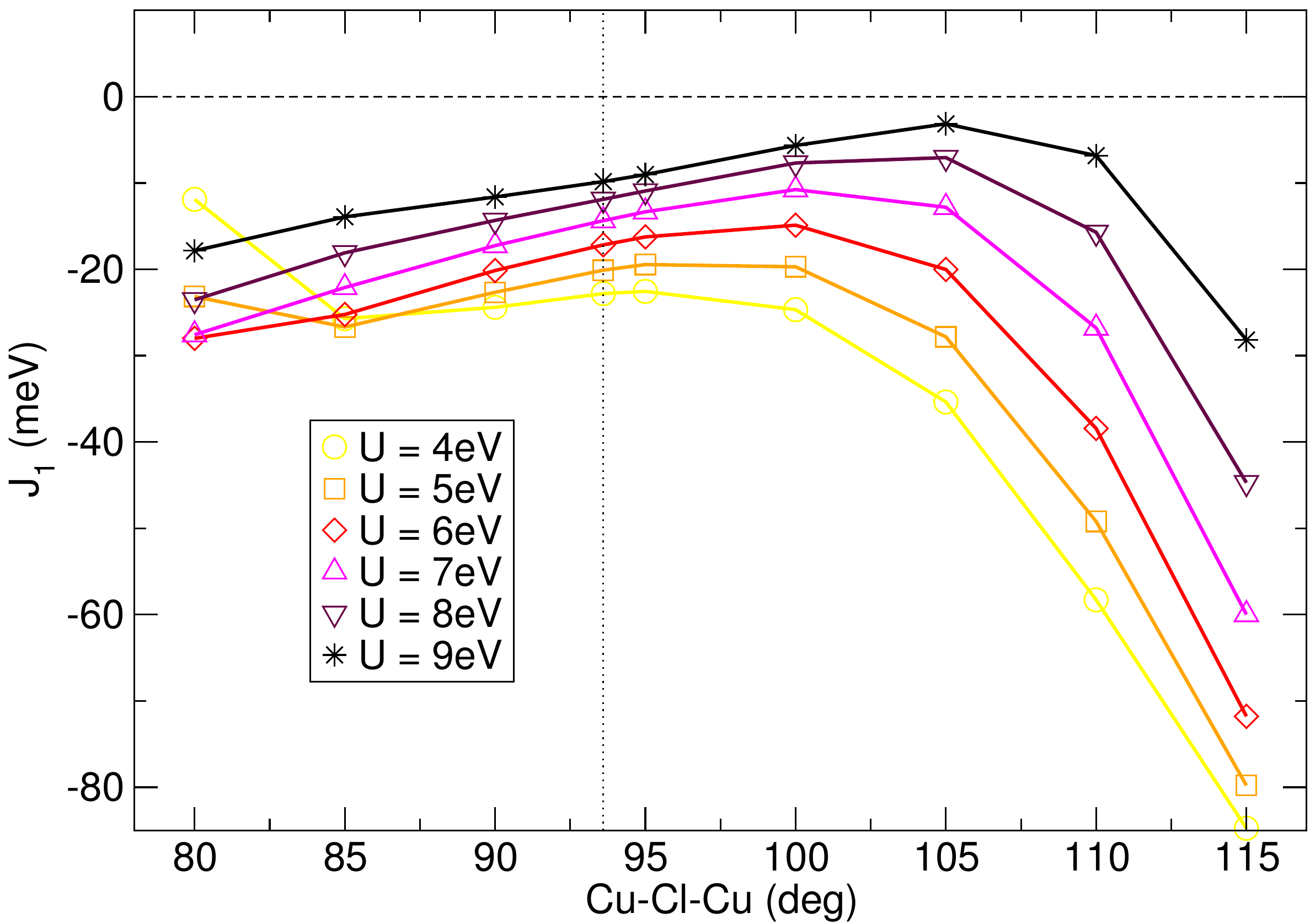}
\begin{flushleft}
(Color online) The nearest-neighbor coupling, $J_1$, of \cucl\ for different Cu-Cl-Cu bridging angles, calculated with LSDA+$U$ where $U_d$ is varied between 4-9\,eV. A negative sign corresponds to ferromagnetic coupling. Varying $U_d$ shifts the curves, however, has no qualitative effect on the magnetic coupling, thus, the ferromagnetic coupling that is obtained even at small and large bridging angles is not an artefact of the choice of $U_d$. 
The decreasing coupling strength for increasing $U_d$ can be attributed to the reduced spatial extension of the Cu(3$d$) orbitals and, thus, the smaller overlap with the Cl(3$p$) orbitals. This entails smaller ligand contributions to the Wannier function which are shown to be proportional to the strength of ferromagnetic coupling. The vertical dotted line indicates the experimental bridging angle.
\end{flushleft}
\end{figure*}

\begin{figure*}[tbp]
\includegraphics[width=9cm]{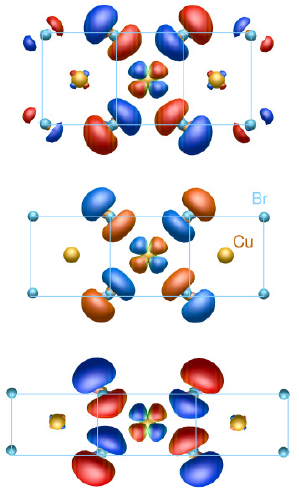}
\begin{flushleft}
(Color online) Wannier functions (WFs) of \cubr\ for bridging angles of 75$^{\circ}$, 90$^{\circ}$ and 115$^{\circ}$. The WFs centered at the Cu-sites are strongly delocalized over the Br-ligands. At 90$^{\circ}$, the small contributions of the neighboring Cu-sites vanish for symmetry reasons, moreover, the ligand contribution is smallest at this bridging angle.
\end{flushleft}
\end{figure*}

\begin{figure*}[tbp]
\includegraphics[width=9cm]{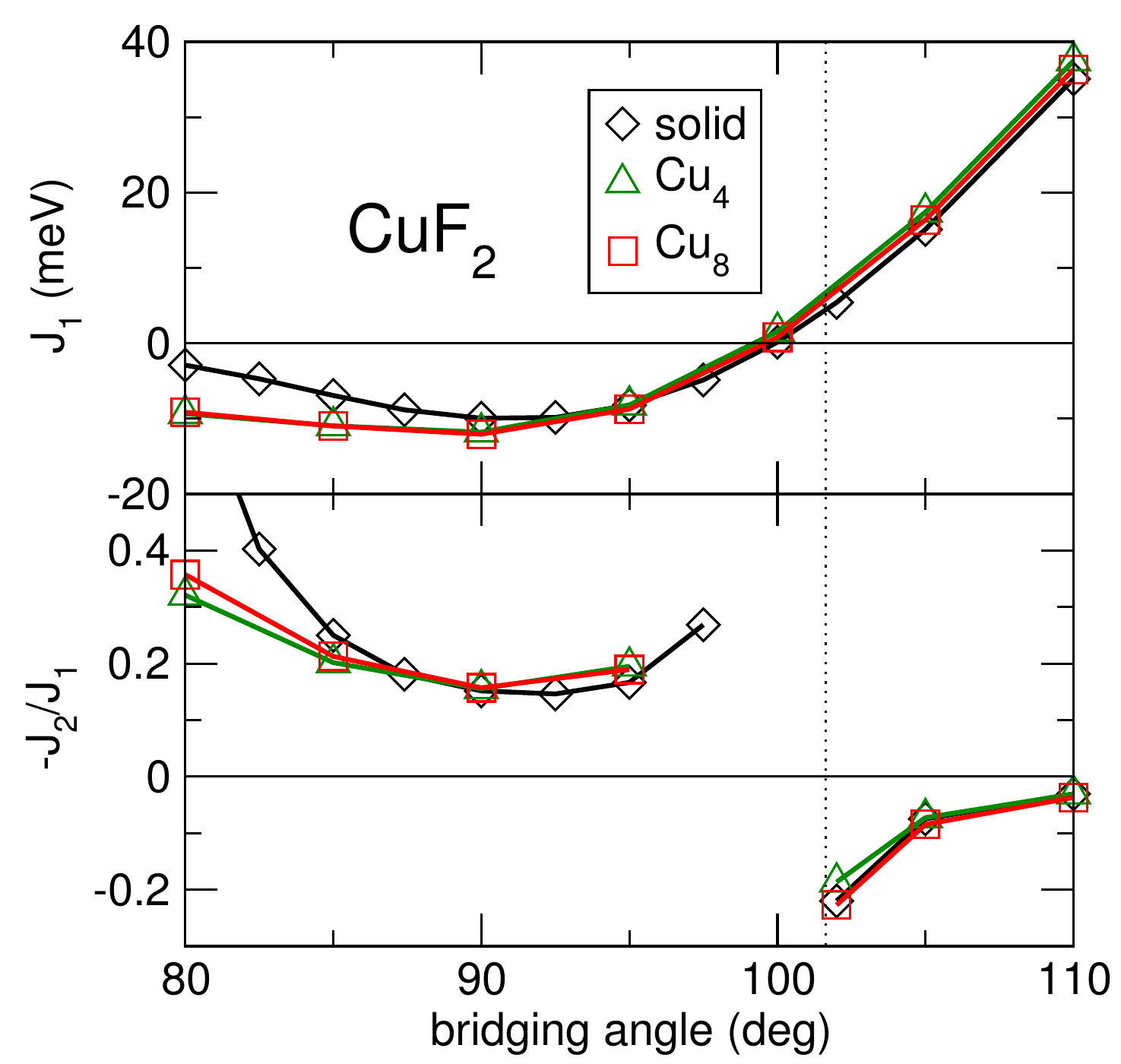}
\begin{flushleft}
(Color online) Magnetic coupling constant $J_1$
and the ratio $\alpha$\,=\,$-J_2/J_1$ of \cuf\ as a function of the
bridging angle calculated with a solid model, Cu$_4$-tetramer and Cu$_8$-octamer cluster models. The dashed vertical line indicates the optimized bridging angle of 102$^{\circ}$. The singularity in the ratio $\alpha$ arises from the change of sign of $J_1$ at about 100$^{\circ}$.
\end{flushleft}
\end{figure*}

\begin{figure*}[tbp]
\includegraphics[width=9cm]{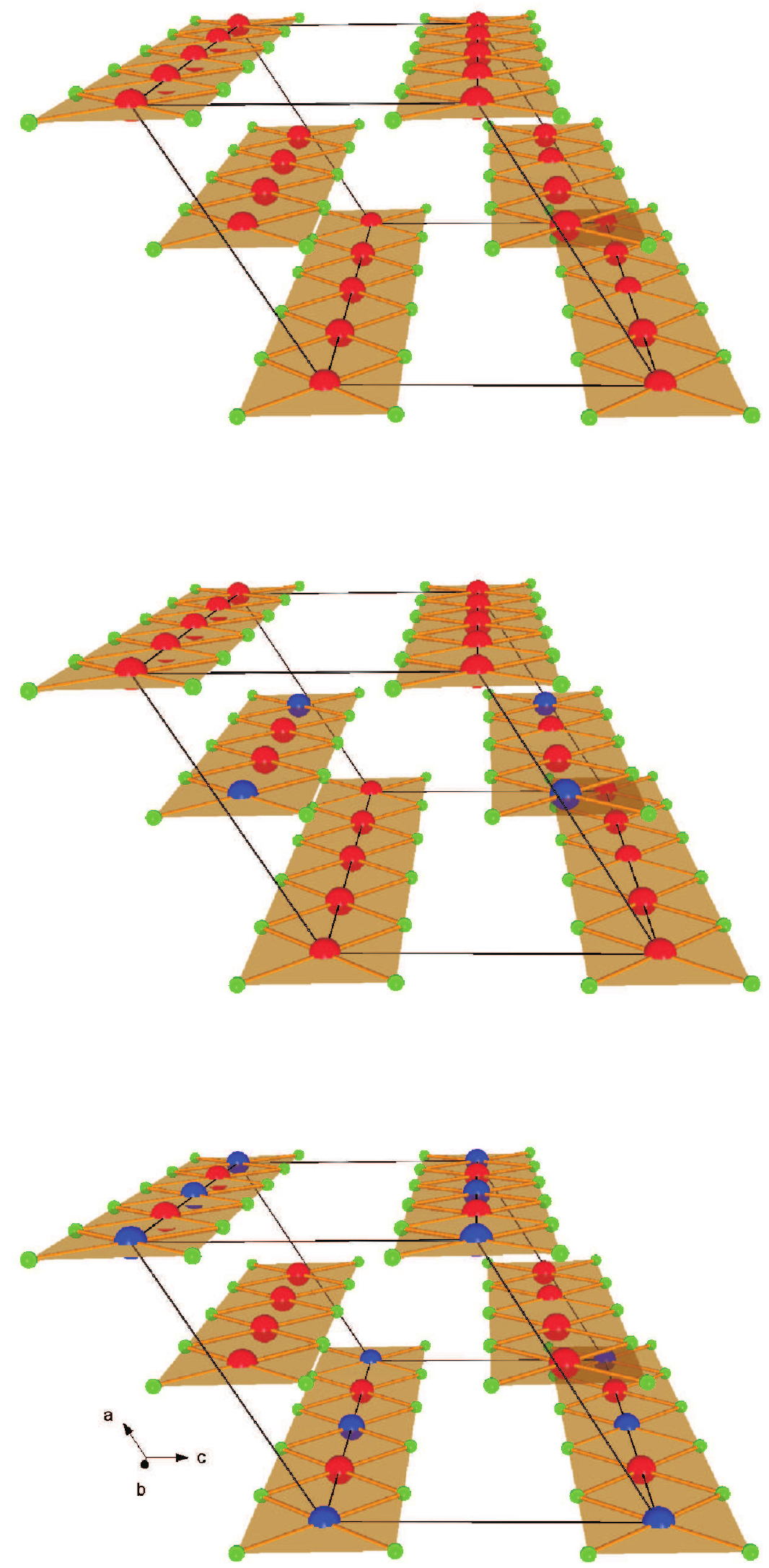}
\begin{flushleft}
(Color online) Supercells quadrupled along the $b$ axis as used for the DFT+$U$ calculations. The total exchange coupling constants $J_1$ and $J_2$ were calculated from total energy differences of these three different arrangements of collinear spins. Red and blue spheres in the structures indicate Cu$^{2+}$ ions with spin up and spin down, respectively. Green balls are F, Cl and Br ligands, respectively.
\end{flushleft}
\end{figure*}

\end{document}